\documentclass[fleqn,usenatbib]{mnras}

\usepackage{newtxtext,newtxmath}

\usepackage[T1]{fontenc}

\DeclareRobustCommand{\VAN}[3]{#2}
\let\VANthebibliography\thebibliography
\def\thebibliography{\DeclareRobustCommand{\VAN}[3]{##3}\VANthebibliography}

\usepackage{graphicx}	
\usepackage{amsmath}	
\usepackage{xcolor,color}
\usepackage{pdflscape}
\usepackage{tabularx}
\usepackage{placeins}
\usepackage{enumitem}

\title[SVOM strategy for the rapid follow-up of GW events]{Optimisation of the SVOM satellite strategy for the rapid follow-up of gravitational wave events}

\author[J. -G. Ducoin et al.]{
J. -G. Ducoin$^{1}$\thanks{E-mail: ducoin@iap.fr},
B. Desoubrie$^{1}$,
F. Daigne$^{1}$,
N. Leroy$^{2}$,
\\
$^{1}$Sorbonne Université, CNRS, UMR 7095, Institut d’Astrophysique de Paris, 98 bis bd Arago, 75014 Paris, France\\
$^{2}$IJCLab, Université Paris-Saclay, CNRS/IN2P3, Orsay, France\\
}

\date{Accepted XXX. Received YYY; in original form ZZZ}

\pubyear{2022}

\begin{document}
\label{firstpage}
\pagerange{\pageref{firstpage}--\pageref{lastpage}}
\maketitle

\begin{abstract}
The SVOM satellite, to be launched in early 2024, is primarily devoted to the multi-wavelength observation of gamma-ray bursts and other higher-energy transients. Thanks to its onboard Microchannel X-ray Telescope and Visible-band Telescope, it is also very well adapted to the electromagnetic follow-up of gravitational wave events. We discuss the SVOM rapid follow-up strategy for gravitational wave trigger candidates provided by LIGO-Virgo-KAGRA. In particular, we make use of recent developments of galaxy catalogs adapted to the horizon of gravitational wave detectors to optimise the chance of counterpart discovery. We also take into account constraints specific to the SVOM platform. Finally, we implement the production of  the SVOM observation plan following a gravitational wave alert and quantify the efficiency of several optimisations introduced in this work. 
\end{abstract}




\section{Introduction}
\label{Introduction}

The first detection of an electromagnetic counterpart of a gravitational wave (GW), 
following the GW signal of 
the binary neutron star (BNS) coalescence of the 17th August 2017 (GW170817), was a real breakthrough for the multi-messenger astronomy \citep{LSC_BNS_2017PhRvL}. It provided the first direct evidence of a link between BNS merger and short gamma ray burst (GRB). The relatively small localisation error (skymap) of this event and the huge effort of 
the community involved in the electromagnetic follow-up allowed the identification of the kilonova counterpart (e.g. \citet{Villar2017,Arcavi2018}) and the afterglow counterpart (e.g. \citet{DAvanzo2018,Hajela2019}), as well as the host galaxy (e.g. \citet{Cantiello2018,Ebrova2020,Yubin2023}). The multi-wavelength observations improved our understanding of many aspects of such extreme
phenomena from the post-merger physics (merger remnant, ejecta, ambient medium), to cosmology (a new independent measurement of the Hubble constant), nuclear physics (neutron star equation of state) 
and fundamental physics (speed of gravitational waves, limits on Lorentz invariance violations in strong-gravity), etc.
\citep[e.g.][]{gwtohubble1,2018PhRvL.121p1101A,Metzgerkilo,2019arXiv190906393H,2019ApJ...876..139G,gwtohubble2,gwtohubble3}.

The third and most recent LIGO-Virgo-KAGRA (LVK) run
(O3) also detected several events of great interest,
including the first confident observations of neutron star-black hole (NSBH) mergers \citep{gwtc2.1,gwtc3}. However, despite a few BNS merger alerts, no new multi-messenger detection occurred. This highlights the challenge of the GW follow-up where one has to deal with large localisation errors from GW detectors and relatively faint and fast decaying transient (afterglow and kilonova). Despite the upgrade of the sensitivity of the GW detectors for the next observing run (O4), the median area of the 90\% credible region is expected to be larger than during O3 because events will be detected at larger distances \citep{O4expect}.

Recent efforts tried to improve the observations of such large regions by optimising the
sky-tiling and the observation plan to cover in a given time the largest possible fraction of the sky localization error box \citep{2016A&A...592A..82G, gwemopt, Coughlin19_opt}. Other developments have taken into account galaxy populations and galaxy properties in the strategy \citep{2016ApJ...820..136G,LosC, Antolini2017, Rana2019,mangrove,glade+}. Such developments 
take advantage of the distance estimation provided by the LVK localisation for each pixel in the skymaps (3D skymaps), but require some understanding of the expected properties of the host galaxies of compact binary mergers. It greatly facilitates the electromagnetic follow-up with narrow field of view (FoV) telescopes as it allows to provide a ranked list of galaxies to be observed in priority. Including priors on host galaxies has raised a recent interest to build galaxy catalogs with a high level of completeness, providing relevant properties including the distance, and covering a volume compatible with the LIGO-Virgo-KAGRA sensitivity \citep{2019ApJ...880....7C,2018MNRAS.479.2374D,mangrove,glade+}.

The SVOM mission is a ground and space-based multi-wavelength observatory aiming at detecting GRBs and other high-energy transient sources \citep{SVOM2016}. This mission is a collaboration between French and Chinese space agencies (CNES and CNSA) and is planned to be launched in early 2024, with a nominal scientific operation phase of 5 years. The scientific objectives of SVOM core program are focussed on GRB studies. The combination of satellite and ground-based instruments will cover the observation of the prompt emission from 4 keV to a few MeV, and even in the visible range in a significant fraction of cases, and of the afterglow from the near-infrared to X-rays. 
In addition, SVOM will benefit from an excellent synergy with other observatories, especially in radio and in high-energy and very-high energy gamma-rays.
The SVOM GRB sample will allow to explore the diversity of the GRB population, including weak/soft nearby events \citep{2020Ap&SS.365..185A} and high-redshift ($z>5$) GRBs, to study the nature of GRB progenitors, the physics of GRB ejecta and associated radiation, and to improve the use of GRBs for cosmology \citep{SVOMwhitepaper}.
The SVOM satellite will be equipped with four instruments: two dedicated to the observation of the GRB prompt emission, a coded-mask gamma-ray imager (ECLAIRs) with a field of view of $2$ sr operating in the 4-150 keV energy range \citep{ECLAIRs}; and a gamma-ray spectrometer (GRM) with a field of view of $5.6$ sr operating in the 15 keV-5 MeV energy range \citep{GRM}. Two telescopes dedicated to the observation of the GRB afterglow after slewing the satellite, a Microchannel X-ray Telescope (MXT)  with a field of view of $64\times64 $ $\textrm{arcmin}^{2}$ operating in the soft X-ray range (0.2-10 keV) \citep{MXT}; and a 40 cm aperture Ritchey-Chrétien Visible-band Telescope (VT) with a field of view of $26\times26$ $\textrm{arcmin}^{2}$ observing in visible (400-650 nm) and in near-infrared (650-950 nm) \citep{VT}. Thanks to its capacity to obtain multi-wavelength follow-up observations, the SVOM mission can play a significant role in the time-domain/multi-messenger astronomy. For this purpose, the time allocated by the SVOM spacecraft to the observation of targets of opportunity (ToO; including GW follow-up) is set to be at least 15\% of the
first two years of scientific operation, and is expected to increase later on. 

\begin{table}
  \hspace{0.5cm}
  \caption{\label{tab:skymap} Properties of the skymaps used to simulate SVOM observation plans. The first 8 lines of the table  correspond to true GW alert skympas and the remaining lines are mock skymaps. D$_\textrm{L}$ is the luminosity distance and its standard deviation reconstructed by bayestar \citep{bayestar}. N$_\mathrm{gal}$ is the number of compatible galaxies as defined in \citet{mangrove}.
  In the tiling strategies studied in this work, we limit ourselves in practice to 2000 galaxies: see text.}
  \begin{tabularx}{\columnwidth}{ c c c c }
    \hline
    \hline
    \rule{0pt}{4ex}
    skymap  & Area  & D$_\textrm{L}$    & N$_\mathrm{gal}$     \\
    name   & [deg$^2$] & [Mpc] &  \\
    \hline\\
    GW170817 & 30 & 40 $\pm$ 8 &  70\\
    GW170817\_no\_Virgo & 190 & 34 $\pm$ 9 &  194\\
    S190425z & 10183 & 155 $\pm$ 45 &  >2000\\
    S190718y & 7246 & 227 $\pm$ 165 &  >2000\\
    S190814bv & 38 & 276 $\pm$ 56 &  821\\
    S190901ap & 13613 & 242 $\pm$ 81 &  >2000\\
    S191213g & 1393 & 195 $\pm$ 59 &  >2000\\
    S200213t & 2587 & 224 $\pm$ 90 &  >2000\\
    MS191219a & 21 & 86 $\pm$ 18 &  49\\
    MS191221a & 828 & 132 $\pm$ 37 &  >2000\\
    MS191221b & 540 & 95 $\pm$ 25 &  >2000\\
    MS191221c & 166 & 61 $\pm$ 15 &  173\\
    MS191222a & 848 & 138 $\pm$ 42 &  >2000\\
    MS191222o & 755 & 119 $\pm$ 30 &  >2000\\
    MS191222t & 555 & 100 $\pm$ 24 &  1086\\
  \end{tabularx}

\end{table}

In this paper we discuss the SVOM rapid follow-up strategy following GW alerts focusing on space instruments. In our simulated observation plan, the optimisations are led by MXT observations. This choice is motivated by the small number of instruments available for the rapid follow-up in X-rays compared to the situation in the visible domain. SVOM should therefore play a significant role in this spectral range. In addition, the  field of view of MXT being about 6 times larger than that of  VT, this choice has a direct impact on the capacity to cover rapidly a large fraction of a typical GW sky-localization error box. We also focus on the search for electromagnetic counterparts (kilonovae and afterglows) and do not discuss the follow-up and characterisation of already identified counterpart candidates. We implemented the galaxy targeting strategy, where the exploration of the sky-localization error-box is based on the location and properties of galaxies within it, and found its efficiency to be better than that of the tilling strategy. The tiling strategy selects a subset of tiles covering the GW error region, from a predetermined set, which covers the entire sky. The observed order of this subset is optimised by the presence of galaxies at a distance consistent with the GW event.
We then improve our simulations by including  constraints specific to the SVOM satellite platform and its onboard detectors, and we implement several optimizations of the galaxy targeting strategy, especial\-ly taking into account the galaxies that are also observable by VT.
This leads to the simulation of observation plans that optimise the chance of counterpart discovery while being realistic about the constraints of the satellite.
Note that this paper studies SVOM rapid follow-up strategy in the case where no GRB has been detected by ECLAIRs on board SVOM in association with the GW alert. In the opposite case, with an associated GRB detected by ECLAIRs, the usual SVOM strategy for the GRB follow-up would apply, based on the accurate localization of the prompt GRB.

In Section \ref{section:Simulation}, we present the simulation methodology. In Section \ref{section:Tillingvsgaltar}, we compare the tilling and the galaxy targeting strategy for SVOM. In Section \ref{section:constraint}, we implement constraints specific to the satellite platform and instruments in the production of the observation plans. In Section \ref{section:furtherdev}, we further improve the galaxy targeting strategy, especially to optimise the follow-up with VT. In Section \ref{section:TBD}, we discuss the
prospects for the GW follow-up by SVOM in the light our results. We conclude in Section \ref{section:conclusion}. Throughout this paper, we use the Plank 2015 cosmological parameters \citep{planck15}.\\

\section{Observation plan simulation}
\label{section:Simulation}

Among the SVOM ToO a significant fraction of them, named ToO\_MM, will be dedicated to the follow-up of multi-messenger alerts. This represents about one ToO\_MM per month. In the current program, up to 24 hours of observation can be allocated to a given ToO\_MM. We focus in this work on ToO\_MM dedicated to BNS merger candidates from LIGO-Virgo-KAGRA as they represent the most promising GW sources of electromagnetic counterparts from gamma-rays to radio. Neutron star-black hole (NSBH) mergers are also promising sources of electromagntic counterparts in cases where the neutron star is tidally destroyed before reaching the black hole horizon. Therefore, some NSBH merger candidates should also be followed, even if GW detections of such events are expected to be rarer \citep{O4expect}.

GW sources represent a challenge because of the size of their skymaps. In the following simulations, we set the exposure time of any image to be 10 minutes which is expected to be a good trade-off between the sensitivity of the MXT telescope to BNS afterglows within the horizon of GW detectors (see Figure~\ref{fig:170817_like}) and the possibility of multiple pointings for the exploration of a large error box. With this exposure time and taking into account the slew maneuver (expected to be lasting less than 5 minutes in the worst case), it allows to have 5 tiles per orbit and a total of 70 tiles for a given ToO i.e. in 24 hours. Throughout this paper we routinely use this number of 70 tiles to quantify the follow-up expectation of the SVOM satellite. The production of the observation plans presented in this paper is one of the early steps of the ToO follow-up system of SVOM. In practice the observation is expected to occur at least few hours after the ToO alert because of the ToO follow-up system procedures following an alert not provided by the SVOM satellite (including human input) with important variation in the delay: the follow-up validation by the French Science Center, the transmission of the plan to the mission center in China, the refinement of the tiling scenario according to the satellite orbit and the communication with the satellite. As a result of this 
reaction delay, we can't take into account the Earth and Moon occultations within this work. On the other and, the Sun is moving sufficiently slowly ($\sim$1 deg per day) to be taken into account in our plan. The SVOM payload constraint imposes the angle between its optical axis and the direction of the Sun to be $>$90 deg. We implement this constraint in the simulation in Section \ref{subsection:sunconstraint}. Another important constraint for the observation of ToO is the limitation of the slew of the satellite. Although the design of the satellite platform has been selected to perform regular and rapid slews, in case of ToO, the angular distance between two different pointings is imposed to be less than 5 degrees within one orbit to prevent an over-stressing of the platform. This constraint is particularly limiting for the follow-up of wide skymap like GW alerts. We discuss in Section \ref{subsection:slewconstraint} the implementation of this slew constraint in the simulation of observation plan.

Within this work we use the \textit{gwemopt}\footnote{\url{https://github.com/mcoughlin/gwemopt}} python package \citep{gwemopt}, developed to optimise the electromagnetic follow-up of GW events and where both tiling and galaxy targeting strategies are implemented. In this work we use the Mangrove galaxy catalogue which cross-matched the GLADE galaxy catalogue with the AllWISE catalogue up to 400 Mpc and derived stellar masses using a mass-to-light ratio using the W1 band luminosity \citep{mangrove}.

In order to simulate a wide variety of observation plans we selected a set of 15 GW skymaps, 8 true alert skymaps (GW170817, GW170817 without Virgo data, S190425z, S190718y, S190814bv, S190901ap, S191213g, S200213t) and 7 mock skymaps (MS191219a, MS191221a, MS191221b, MS191221c, MS191222a, MS191222o, MS191222t) published by LIGO-Virgo-KAGRA through the GW Candidate Event Database (GraceDB\footnote{\url{https://gracedb.ligo.org/}}). All of them are selected to be with a mean distance plus standard deviation below 400 Mpc allowing to use the Mangrove catalog. They are also chosen to represent the variety of sky localisations provided by the GW detectors alone, with skymaps representative of a detection with 1, 2 or 3 detectors. Table \ref{tab:skymap} presents the properties of these skymaps. The observation plans produced with the \texttt{gwemopt} \citep{gwemopt} software are adapted to ground based observatories and not to a satellite. Therefore, we implemented within this work the tools necessary to produce observation plans for the SVOM satellite. This includes getting rid of the ground observations limitations (horizon, azimuth...) and the addition of limitations required by SVOM observations, as discussed in Section \ref{section:constraint}.

\begin{figure}
\begin{center}
\includegraphics[width=0.7\columnwidth]{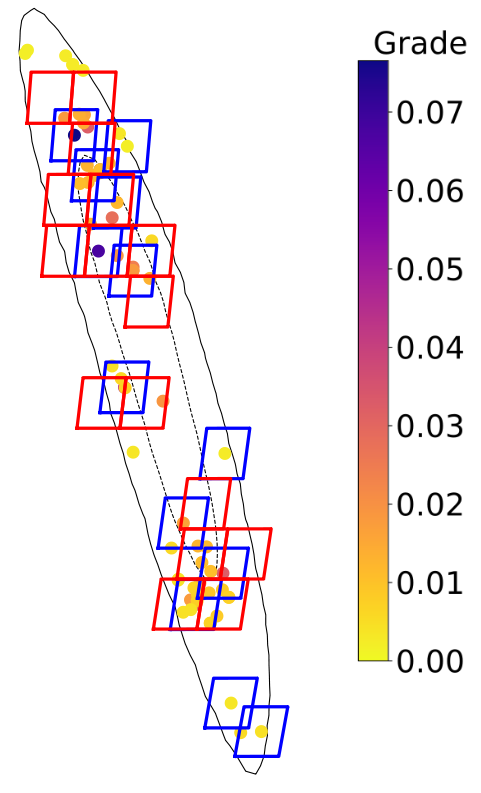}
\caption{Skymap of GW170817. The dotted and the solid line enclose the regions of the skymap with respectively a 50\% and 90\% probability of presence for the GW source. The dots show the location of the compatible galaxies and their color represent their grade i.e. the quantity used to rank galaxies by favoring those whose properties make them the most credible host candidates (see Section \ref{section:Tillingvsgaltar}). The 16 red (resp. 13 blue) squares show all the tiles of the observing plan obtained  with the galaxy-weighted tiling strategy (resp. the galaxy targeting strategy).}
\label{fig:obtained_tiles}
\end{center}
\end{figure}

\section{Tiling vs Galaxy targeting}
\label{section:Tillingvsgaltar}
As explained in Section~\ref{Introduction}, SVOM observation plans for the rapid follow-up of GW alerts are led by the X-ray telescope (MXT), mainly because of its field of view (the consideration of VT observations is discussed in Section \ref{section:furtherdev}). In this section we focus on the following question: should the MXT telescope use the tiling strategy or the galaxy targeting strategy for its observations? 

Both strategies take advantage of galaxy catalogs to optimise the observations. This idea starts from the hypothesis that the source is located within (or nearby) a galaxy, the host galaxy of the BNS system, as commonly observed in the case of short GRBs \citep{2017ApJ...848L..13A, 2022arXiv220601763F} or was the case for GW170817 within the galaxy NGC4993 (e.g. \citet{Cantiello2018,Ebrova2020,Yubin2023}). For compact binary coalescence, the HEALPix skymap provided with the GW alert (HEALPix: Hierarchical Equal Area isoLatitude Pixelization, \citealt{HEALPix}) also provides the estimated distance of the source. For each pixel of the skymap, one can fetch the probability distribution for the source distance at the given sky position of the pixel. Hence, one can select galaxies compatible with such a 3D skymap. We classified as "compatible" with a given skymap, a galaxy which fulfills the two following conditions:
\begin{enumerate}
    \item its 2D position in the sky has to be in the $90\%$ of the 2D skymap probability distribution;
    \item its distance has to fall within the 3 sigma distance error localization at the given pixel of the galaxy.
\end{enumerate}
Further development also leads to assign a grade to each compatible galaxy according to some galaxy properties \citep{LosC,mangrove}: in this work we use the definition of the grade given in equation (4) of \citet{mangrove}, which is based mainly on the stellar mass of the galaxy. This is motivated by several works pointing out a significant dependence to the stellar mass for the rate of BNS merger \citep{2019MNRAS.487.1675A, 2019MNRAS.tmp.2085T, 2018MNRAS.481.5324M, 2022arXiv220505099S} and the population of massive short GRB host galaxies 
\citep{Leibler2010, Fong2013, Berger2014, 2022arXiv220601764N}.
We then rank compatible galaxies according to their grade, keeping only the first 2000 ones if they are more numerous.
This number is chosen to ensure that only galaxies with very low grade are removed and 
that the observation plan includes at least 70 tiles so that all the observation time allocated to the Too\_MM is used.

\begin{table*}
  \hspace{0.5cm}
  \caption{Description of the naming convention used throughout this work. We consider an observation plan with a total number of $N_{\mathrm{tiles,max}}$ tiles.
  The index $i$ runs over tiles and the index $j$ over  galaxies contained in a given tile and compatible with the considered 3D-skymap.
  The grade of each compatible galaxy is written $G_{\mathrm{gal,}j}$ (see text).
 At an intermediate stage of the implementation of the observation plan, the number of tiles already observed is $N_{\mathrm{tiles}}\le N_{\mathrm{tiles,max}}$.
  }
  \label{tab:naming}
  \begin{tabular}{ c c c}
    \hline
    Naming & Description  & Expression\\
    \hline
    Number of galaxies of a tile & Number of compatible galaxies contained by tile $i$ & $N_{\mathrm{gal,} i}$\\
    Grade of a tile & Sum of the grades of the compatible galaxies contained by the tile $i$ & $\sum_{j = 0}^{N_{\mathrm{gal,}i}} G_{\mathrm{gal,}j}$ \\
    \hline
    Number of observed galaxies & Number of galaxies observed within $N_{\mathrm{tiles}}$
    & $\mathcal{N}_{\mathrm{gal}}=\sum_{i = 0}^{N_{\mathrm{tiles}}} N_{\mathrm{gal},i}$\\
    Observed grade
    & Sum of the grades of the $N_{\mathrm{tiles}}$ observed
    & $\mathcal{G}=\sum_{i = 0}^{N_{\mathrm{tiles}}} \sum_{j = 0}^{N_{\mathrm{gal},i}} G_{\mathrm{gal},j}$\\
    \hline
    Maximum number of observed galaxies & 
    Sum of the number of galaxies of all tiles in the observation plan
    & $\mathcal{N}_{\mathrm{gal,max}}=\sum_{i = 0}^{N_{\mathrm{tiles,max}}} N_{\mathrm{gal},i}$ \\
    Maximum 
    observed grade  & 
    Sum of the grades of all the tiles in the observation plan
    & $\mathcal{G}_{\mathrm{max}}=\sum_{i = 0}^{N_{\mathrm{tiles,max}}} \sum_{j = 0}^{N_{\mathrm{gal},i}} G_{\mathrm{gal},j}$ \\
    \hline
  \end{tabular}
\end{table*}

In the following, regardless of the strategy, we define the \textit{number of galaxies of a tile} as the number of compatible galaxies it contains, we define the \textit{grade of a tile} as the sum of the grades of the compatible galaxies it contains. 
A given \textit{observation plan} is an ordered sequence of $N_{\mathrm{tiles,max}}$ tiles to observe. When $N_{\mathrm{tiles}}\le N_{\mathrm{tiles,max}}$ have already been observed, we define the \textit{number of observed galaxies} (resp. the \textit{observed grade}) as the sum of the number of galaxies (resp. the sum of the grades) of all the already observed tiles. If all tiles of the observation plan have been observed, the final value of the number of observed galaxies (resp. of the observed grade) is called the \textit{maximum number of observed galaxies} (resp. the \textit{maximum observed grade}). 
For clarity, these definitions are summarized in Table \ref{tab:naming}.

\subsection{Tiling strategy}
\label{subsection:tiling}

The tiling strategy usually used by large FoV ($\gtrsim  1$ deg$^{2}$) telescopes consists first of the construction of an optimised tiling of the sky and then of the scheduling of the observation of these tiles based on the localisation (probability distribution) from the considered GW skymap: (1) 
The entire sky is divided into tiles. This is performed once and in advance of any GW follow-up.
In this strategy, the size of the tiles is defined to fit the telescope field of view so that a tile corresponds typically to a single image. The optimized tiling of the sky is built in such a way that the overlap between the tiles is minimized to observe the largest sky area possible with a given number of tiles; (2) then, for a given GW alert, the  scheduling of the observation is based on the ranking of these tiles according to the probability to host the GW source. This probablity is computed for each tile by summing over all pixels of the tile
the probability for the GW source to be in the sky direction of the pixel. This 2D probability is provided by the GW skymaps which consists in all-sky pxelised images using the HEALPix format.
A further extension of this strategy is to incorporate galaxy catalogues to focus the observations on compatible galaxies for a given GW alert.
We call this extension the \textit{galaxy-weighted tiling strategy}.
In this strategy, rather than using the probability to host the GW source, we rank the tiles according to their grades, as defined above and given in Table~\ref{tab:naming}.

This galaxy-weighted tiling strategy (as well as the galaxy targeting strategy presented in the next subsection) avoids to observe regions of the sky were there is no compatible galaxy. This well illustrated in Figure~\ref{fig:obtained_tiles}, where a significant fraction of the GW170817 skymap is not observed in our observation plan due to the lack of compatible galaxies. The standard tiling strategy is not considered in this work. We rather focus on the galaxy-weighted tiling strategy, which is compared to the galaxy targeting strategy described in the next subsection.

\subsection{Galaxy targeting strategy}
\label{subsection:galaxytargeting}

The \textit{galaxy targeting strategy} is usually used for small FoV telescopes.
In this strategy, a new optimized tiling of the sky is built for each GW alert, based on the galaxies compatible with the GW skymap. Once all these galaxies are ranked according to their grade (see above), the first tile (still with a size corresponding to the instrument FoV) is centered on the galaxy ranked first. This galaxy and all other galaxies contained in this tile are removed from the list. The second tile is centered on the new galaxy ranked first in the remaining list, and this procedure is repeated until the list is completely exhausted.

As illustrated in Figure \ref{fig:obtained_tiles}, the galaxy targeting strategy is more flexible than the galaxy-weighted tiling strategy and often reduces the number of pointings necessary to observe a list of galaxies. On the other hand, it may lead to more overlap between tiles than the galaxy-weighted tiling strategy which is precisely built to limit this overlap. The benefit of one strategy over the other mainly depends on the telescope FoV but also on the GW skymap size, shape and distance. In particular, the MXT FoV ($\sim$ 1 $\textrm{deg}^2$) is in a range where choice between the two strategies is not obvious.

\begin{figure*}
\begin{center}
\centering
\includegraphics[width=2.1\columnwidth]{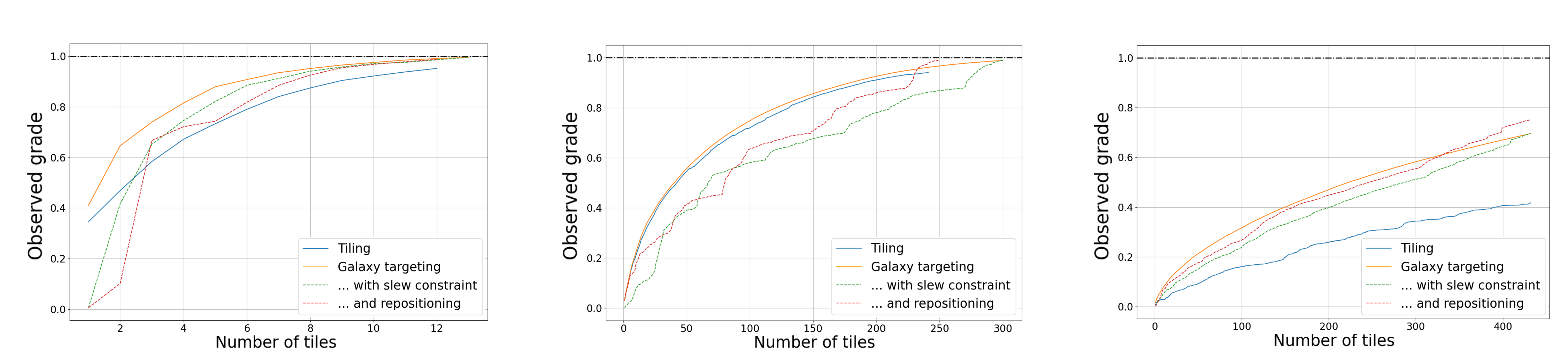}
\end{center}
\caption{
Comparison of the evolution of observed grade $\mathcal{G}$ normalized by the maximum observed grade $\mathcal{G}_{\mathrm{max}}$ (see Table~\ref{tab:naming} for definitions) as a function of the number of tiles already observed $N_{\mathrm{tiles}}$ for simulated observing plans for the rapid follow-up of three representative GW skymaps corresponding to a detection with 3 (GW170817), 2 (MS19122t) and 1 (S190901ap) detector using either the galaxy-weighted tiling strategy discussed in \S~\ref{subsection:tiling} or the galaxy targeting strategy either in its first version discussed in \S~\ref{subsection:galaxytargeting}, or improved by taking into account the slew constraint of SVOM discussed in \S~\ref{subsection:slewconstraint}, or even further optimised with the repositing procedure discussed in Section~\ref{section:furtherdev}.
}
\label{fig:three_skymaps_observed_grade}
\end{figure*}

\subsection{Comparison}

We compare the efficiency of both strategies by looking for each skymap at the evolution of the number of observed galaxies $\mathcal{N}_{\mathrm{gal}}$ and the observed grade $\mathcal{G}$ as the observation plan is implemented. The result is shown in Figure~\ref{fig:three_skymaps_observed_grade} for three  skymaps (S190901ap, MS191222t and
GW170817) representative of the localization achieved for a detection by respectively 1, 2 or 3 GW detectors.

\begin{figure*}
\begin{center}
\begin{tabular}{c}
\includegraphics[width=0.9\linewidth]{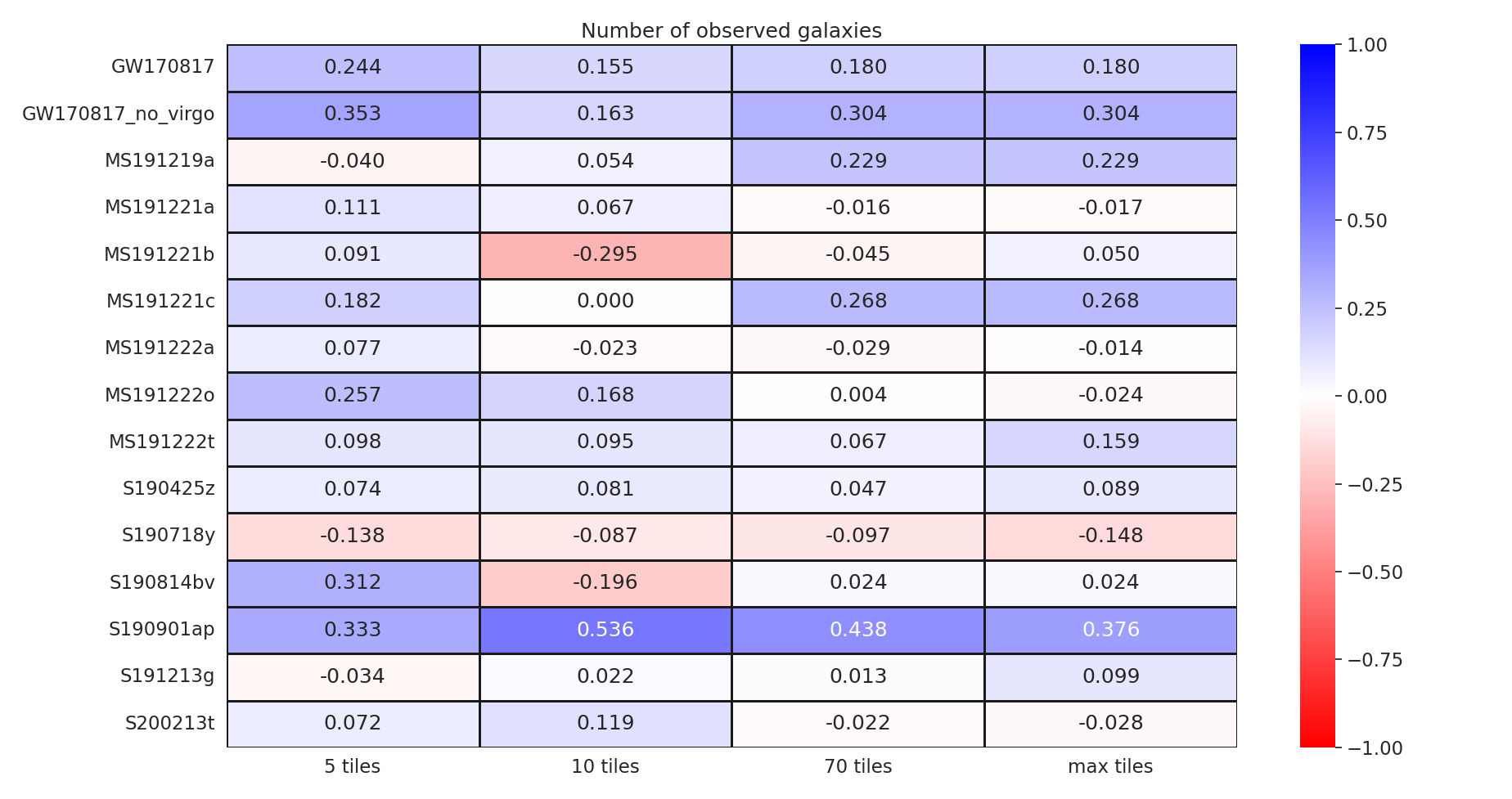}\\
\includegraphics[width=0.9\textwidth]{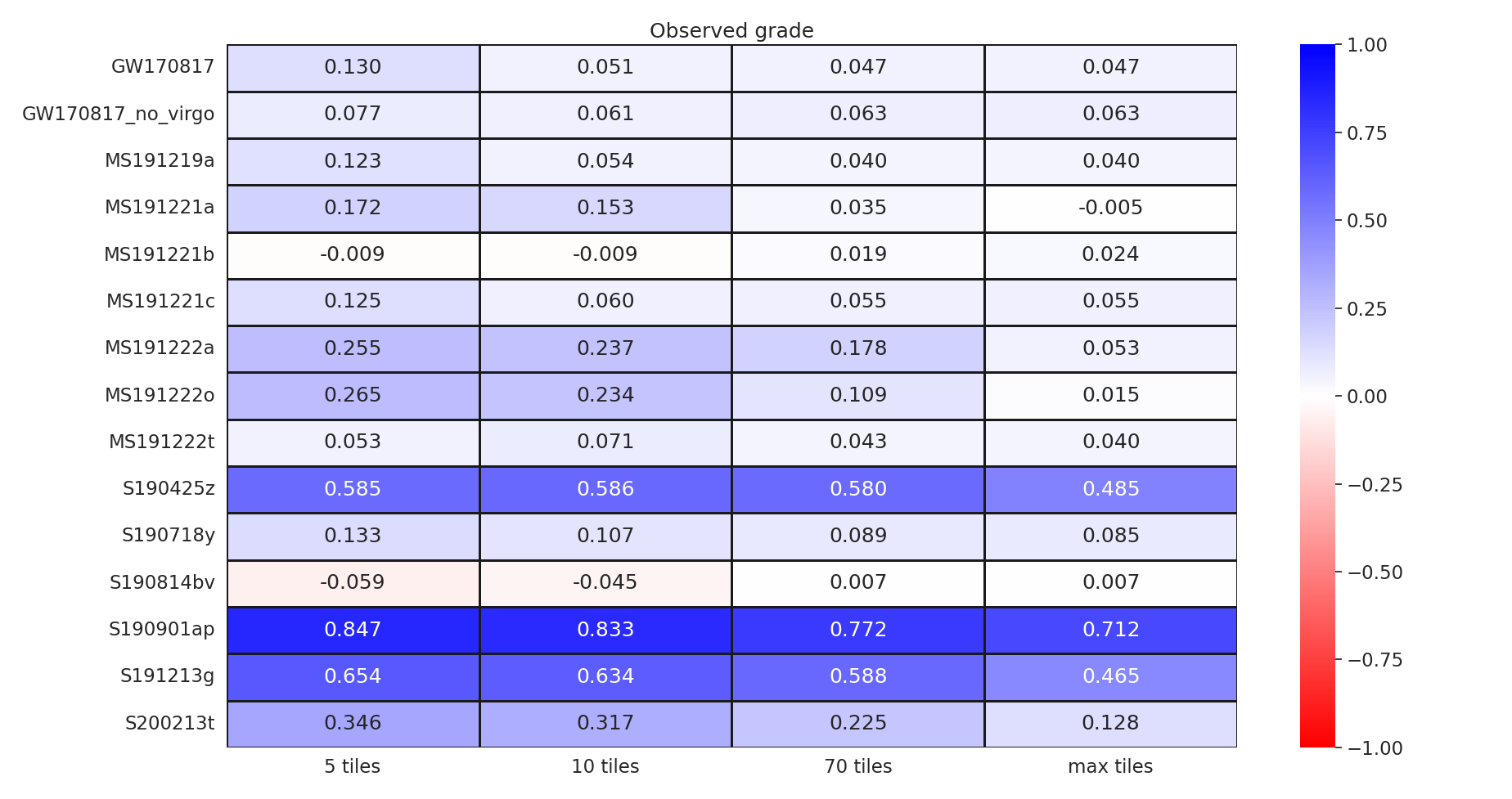}\\
\end{tabular}
\end{center}
\caption{\textbf{Comparison between the tiling and the galaxy targeting strategies:} number of observed galaxies and observed grade. The difference between the number of observed galaxies $\mathcal{N}_{\mathrm{gal}}$ normalized by the maximum number of observed galaxies $\mathcal{N}_{\mathrm{gal,max}}$ (top) or between the observed grade $\mathcal{G}$ normalized by the maximum observed grade $\mathcal{G}_{\mathrm{max}}$ (bottom) with the galaxy targeting strategy and the galaxy-weighted tiling strategy is given for each skymap listed in Table~\ref{tab:skymap}, assuming that  $N_{\mathrm{tiles}}=5$, $10$, $70$ and $N_{\mathrm{tiles,max}}$ tiles have already been observed. The bluer the color, the better the galaxy targeting strategy compared to the galaxy-weighted tiling strategy. Note that 70 tiles is the expected number of tiles allocated for a given ToO with SVOM.}
\label{fig:heatmap_gal_grade}
\end{figure*}

The comparison between the two strategies is summarized in figures
~\ref{fig:heatmap_gal_grade}, which gives the difference between the number of observed galaxies $\mathcal{N}_{\mathrm{gal}}$ (top) and the observed grade $\mathcal{G}$ (bottom) with the galaxy targeting strategy and the galaxy-weighted tiling strategy, when respectively $N_{\mathrm{tiles}}=5$, $10$, $70$ and $N_{\mathrm{tiles,max}}$ tiles have already been observed. This difference is normalized by the maximum number of observed galaxies $\mathcal{N}_{\mathrm{gal,max}}$ (resp. the maximum observed grade $\mathcal{G}_{\mathrm{max}}$), which would be reached if all tiles of the observation plan were observed. 
In one hand, there is no clear evidence in the top panel for an advantage for either strategy
in terms of number of observed galaxies. On the other hand, the bottom panel clearly shows the optimisation of the galaxy targeting strategy in terms of observed grade. This advantage is decisive as the observed grade has been introduced to quantify how likely one will observe the host galaxy of the GW source.

In conclusion these simulations show that the MXT FoV is still in a range where it can benefit from the galaxy targeting approach for the follow-up of gravitational waves.
In addition, the computation time for the galaxy targeting strategy (always less than 1 minute) is significantly smaller than the one for tilling strategy (typically 15 minutes). This could be 
an advantage in the context of the rapid follow-up of a GW alert.
In light of these results, the rest of this study is devoted only to the galaxy targeting strategy, which is developed in a more realistic way by taking into account additional constraints.

\section{Constraints of the satellite}
\label{section:constraint}

In this section, we improve the production of observation plans in the galaxy targeting strategy by taking into account further constraints imposed by the satellite and its platform. In order to quantify the cost of these constraints, we compare for each addition the resulting simulated observation plan with the initial plan produced with the first version of the galaxy targeting strategy described in Section \ref{section:Tillingvsgaltar}. For concision, we limit some figures to the results obtained for three GW skymaps.

\subsection{Sun constraint}
\label{subsection:sunconstraint}
\begin{figure*}
\begin{center}
\includegraphics[width=\textwidth]{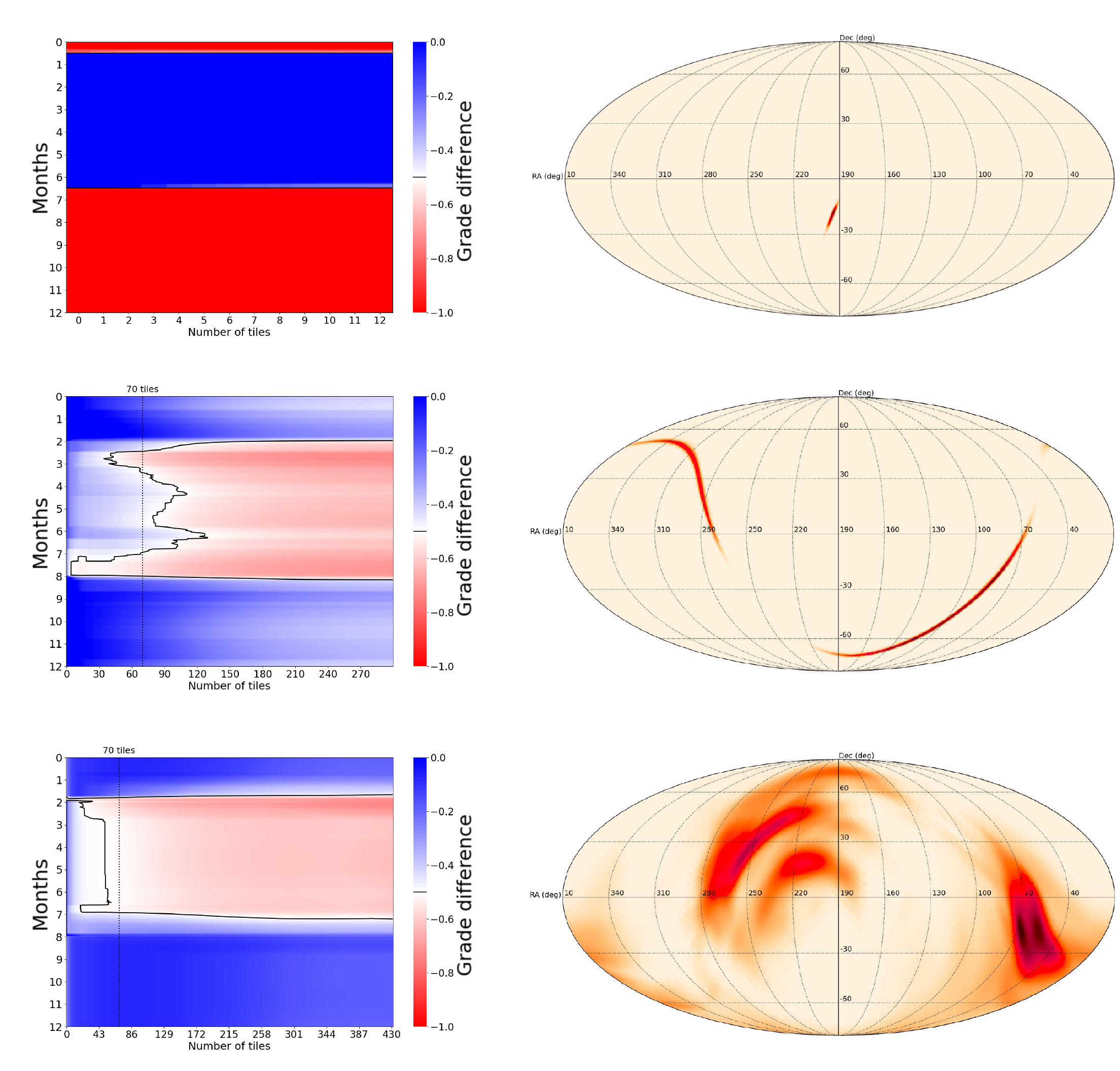}
\caption{
\textbf{Galaxy targeting strategy: impact of the Sun constraint.}
\textit{Left}: Difference between the observed grade G, obtained without and with the Sun constraint, as a function of the number of already observed tiles $N_\mathrm{tiles}$ with the observation plan described in \S~\ref{subsection:sunconstraint} 
for 365 different alert time (one per day). The black line represents the number of tiles $N_\mathrm{tiles}$ at which the constrained observation plan reaches only 50\% of the normalized observed grade of the unconstrained plan. The vertical dashed line highlights the $N_{\mathrm{tiles}}=70$ tiles scenario expected for SVOM. \textit{Right}: Skymaps of GW170817, MS191222t and S190901ap (from top to bottom). The color represents the 2D probability of presence of the host, the darkest color corresponding to the most probable region.}
\label{fig:sun_constraint}
\end{center}
\end{figure*}

One of the main constraints for ToO observations are the Sun occultations. We implemented this constraint within \textit{gwemopt} by imposing any pointing to be at least $>$91 deg away from the sun (one degree more than the system constraint to take into account the reaction delay discussed in Section \ref{section:Simulation}). In order to quantify the impact of this constraint for the follow-up of GW alerts, which usually show very specific skymap shapes, we compare in Figure~\ref{fig:sun_constraint} the observation plan obtained for 3 GW skymaps using 365 different alert time (one per day) with and without including the Sun constraint. These 3 skymaps (S190901ap, MS191222t and  GW170817) are representative of the localization achieved with a detection by respectively 1, 2 or 3 GW detectors. 

Figure~\ref{fig:sun_constraint} shows that in the case of a well localised event such as GW170817, the Sun constraint is similar to what one can expect for a point-like target with a $>$90 deg constraint: roughly, the skymap is entirely observable half of the time and not observable the other half of the time. 

On the other hand, for a larger skymap representative of a detection with two detectors such as MS191222t, the Sun constraint is less limiting. Indeed, in this example, the constrained observation plan reaches less than 50\% of the normalized grade of the unconstrained plan during only 2 months in the $N_{\mathrm{tiles}}=70$ scenario expected for SVOM.
This can be intuitively understood looking at the shape of the skymap and the presence of two distinct high probability regions, typical of this kind of skymap, that are unlikely to be affected both at the same time by the Sun constraint. 

Finally, in a case  representative of a detection with a single detector such as S190901ap, the Sun constraint significantly affects the observation plan for about 5 months in the year.
This highlights the difficulty of the follow-up with such a very large skymap.

\subsection{Slew constraint}
\label{subsection:slewconstraint}

\begin{figure}
\begin{center}
\includegraphics[width=1\columnwidth]{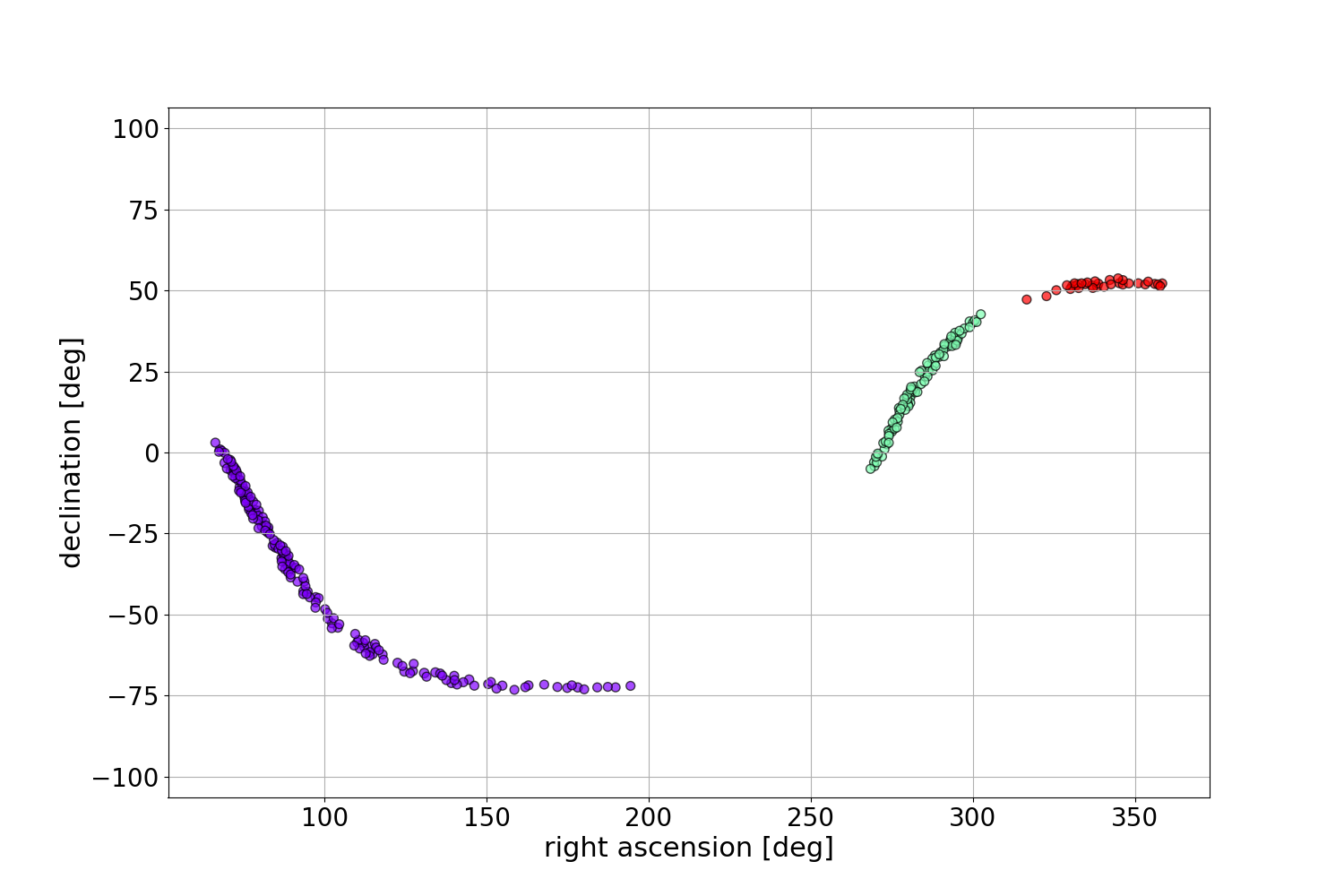}
\caption{Identification of three clusters of tiles in the observation plan generated for MS191222t using the galaxy targeting strategy. Each color corresponds to a different cluster.}
\label{fig:cluster_MS191222t}
\end{center}
\end{figure}

We included the slew constraint on the production of observation plans by adding a post-processing after the \textit{gwemopt} computation. The idea is to re-order 
the sequence of tiles in the obtained observation plan to limit the number of slews by more than 5 degrees. For this purpose, we start from the observation plan generated with the galaxy targeting strategy presented in \S~\ref{subsection:galaxytargeting} and we identify clusters of tiles using a DBSCAN (density-based spatial clustering of applications with noise) algorithm. An example of the obtained clustering is presented in Figure~\ref{fig:cluster_MS191222t} for MS191222t. The clusters are defined so that the tiles in each cluster can be ordered in a sequence such that none of the slews are larger than 5 deg. However, this optimal sequence often requires observations to start at one edge of the cluster. 
In our case this is sub-optimal as the most probable regions for GW skymap are usually at the center of a cluster and these most probable regions are expected to be observed as fast as possible. In order to find an optimal trade-off between the number of slews by more than 5 degrees and the rapid observation of the most probable regions, we define the following procedure to re-order the sequence of tiles for a given cluster:

\begin{enumerate}
    \item Identify
    the barycenter of the tiles of the cluster, defined as the barycenter of their positions weighted by their grades.
    
    \item Select arbitrary the first tile.  
    
    \item Select the next tile as follows:
    \begin{itemize}
        \item If there are remaining tiles that are observable respecting the slew constraint,
        we select the one which minimise the ratio of its distance to the barycenter over its grade.
    
        \item If there are no more tiles that are observable respecting the slew constraint, we select the remaining tile with the highest grade (regardless of its distance to the barycenter).
    \end{itemize}
    \item Repeat step (iii) until all tiles are inserted in this sequence.
\end{enumerate}

We store this re-ordered sequence of tiles of the cluster, the corresponding number of slews by more than 5 degrees, and the corresponding evolution of the observed grade $\mathcal{G}$ as a function of the number of already observed tiles $N_{\mathrm{tiles}}$ as the sequence is implemented. We repeat this procedure by selecting successively each tile of the cluster as the first tile at step (ii). We finally select in our final observation plan the ordered sequence for each cluster that minimise the number of slews by more than 5 degrees. If there is more than one possible sequence for a given cluster, we select the one optimising the evolution of the observed grade as the sequence is implemented.

\begin{figure}
\begin{center}
\includegraphics[width=1\columnwidth]{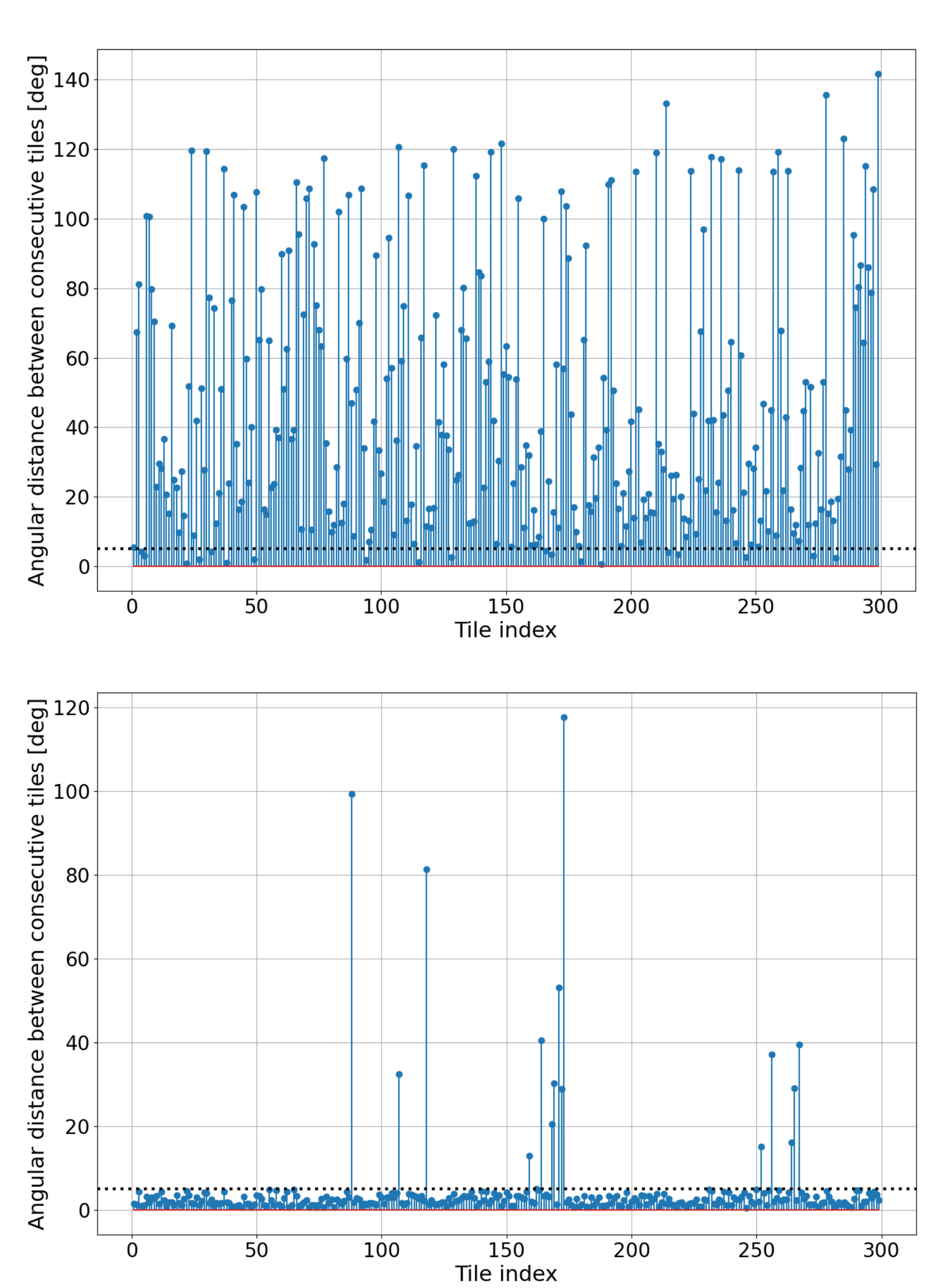}
\caption{\textbf{Galaxy-targeting strategy: impact of the slew constraint.} \textit{Top}: Angular distance between consecutive tiles in the ordered sequence of the observation plan generated for MS191222t with the galaxy targeting strategy, without implementing the slew constraint discussed in \S~\ref{subsection:slewconstraint}. The dotted horizontal line highlights the 5 degrees constraint. \textit{Bottom}: Same figure after the reordering of the 
sequence of tiles defined in \S~\ref{subsection:slewconstraint} to limit the number of slews by more than 5 degrees.}
\label{fig:slew_limitation}
\end{center}
\end{figure}

Figure \ref{fig:slew_limitation} present the angular distance between consecutive tiles obtained with the galaxy targeting strategy before and after the reordering of the sequence of tiles in each cluster according to this procedure. This figure shows that the implemented post-processing is efficient, as it has strongly limited the number of slews by more than 5 degrees. Note that keeping a few slews by more than 5 degrees in the case of a large skymap as for MS191222t  is inevitable, if only to jump from a cluster to another. The few large slews still present in the observation plan are flagged. 

In practice, it is envisaged for SVOM to wait for the South-Atlantic Anomaly (SAA) passages (where the detectors are temporally turned off and after which it is needed to re-point) to perform such large slews. The averaged exposure loss due to the SAA is estimated to be about 18\%. In addition, including the Sun constraint discussed in \S~\ref{subsection:sunconstraint} will also help to limit the number of large slews as it usually makes some of the clusters impossible to observe. 

\begin{figure*}
\begin{center}
\includegraphics[width=0.9\textwidth]{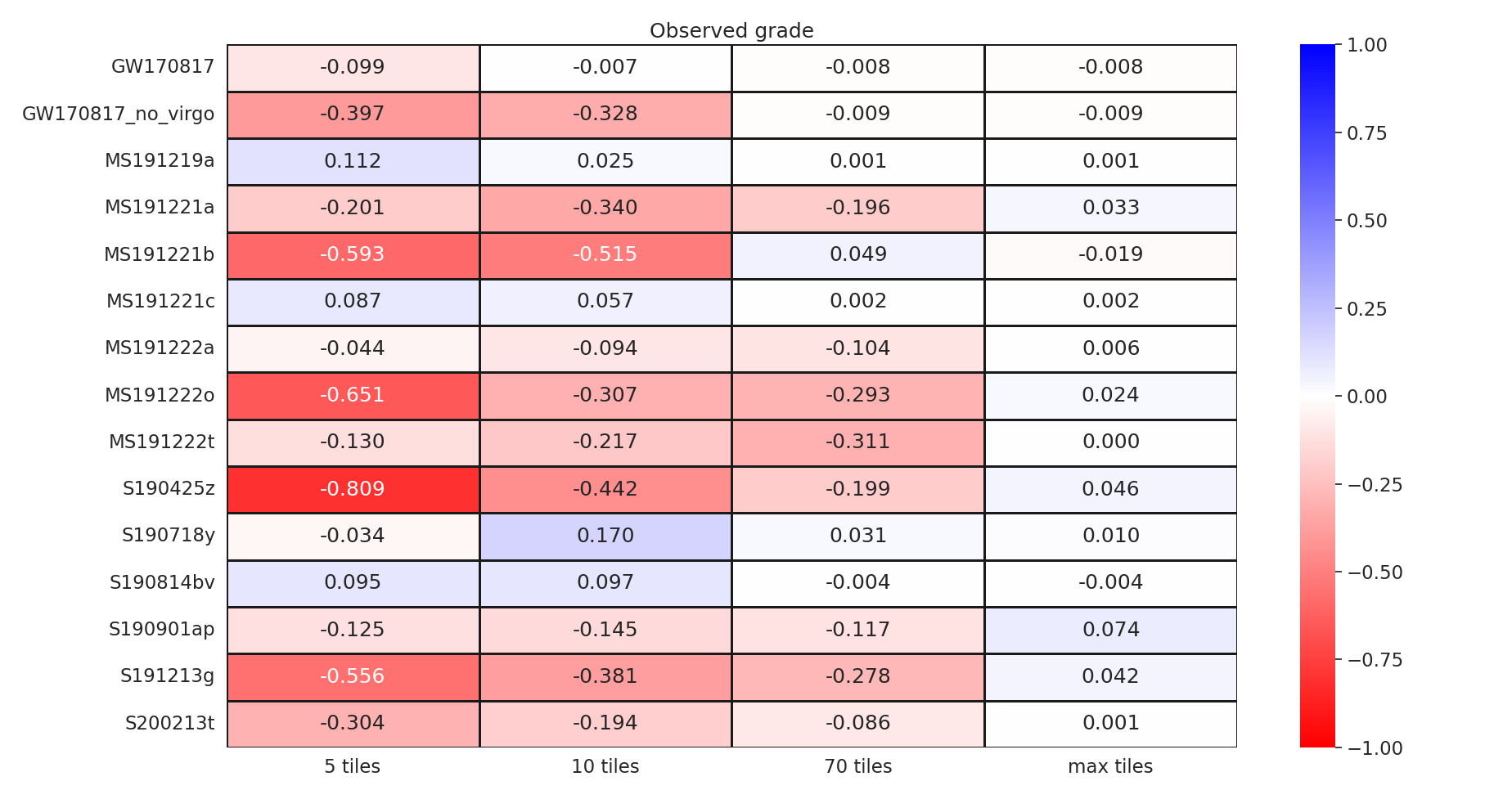}
\caption{\textbf{Comparison between the galaxy targeting strategy with and without including the slew constraint: observed grade.}
The difference 
between the observed grade $\mathcal{G}$
with the galaxy targeting strategy including the slew constraint (following the procedure defined in \S~\ref{subsection:slewconstraint} and without including it
is given for each skymap listed in Table~\ref{tab:skymap}, assuming that  $N_{\mathrm{tiles}}=5$, $10$, $70$ and $N_{\mathrm{tiles,max}}$ tiles have already been observed.
This difference is normalized by the maximum observed grade $\mathcal{G}_{\mathrm{max}}$ (see Table~\ref{tab:naming} for the definition of these quantities).
The bluer the color, the better the galaxy targeting strategy implementing the slew constraint. Note that 70 tiles is the expected number of tiles allocated for a given ToO with SVOM.}
\label{fig:heatmap_grade_galaxy_all_constraint}
\end{center}
\end{figure*}

The implementation of this very restrictive slew constraint has an impact on the expected efficiency of the SVOM rapid follow-up of GW alerts. Figure~\ref{fig:heatmap_grade_galaxy_all_constraint} illustrates the impact of this constraint looking at the observed grade with and without the constraint. It shows that the impact of the constraint tends to be maximal for the few first tiles and decreases with the number of pointings until becoming very small at 70 tiles. Figure~\ref{fig:three_skymaps_observed_grade} also illustrates this comparison by plotting the evolution of the observed grade as a function of the number of already observed tiles. 

We note to conclude that this slew constraint could be slightly relaxed in further developments by strictly complying with the system requirement that imposes a maximum of one slew by more that 5 degress per orbit. However such this optimization cannot be easily simulated as its implementation in the observation plan likely requires the information of the satellite position in its orbit. 

\section{Further development of the galaxy targeting strategy}
\label{section:furtherdev}

\subsection{Tile repositioning procedure}
\label{subsection:reposition}

In the galaxy targeting strategy, the tiles are centered on galaxies 
compatible with the skymap, as defined in Section~\ref{section:Tillingvsgaltar}. This procedure 
can be seen as too restrictive as it is not necessary for a galaxy to be right in the middle of an image for a proper detection (edge conditions are taken into account in the following). For this reason, we present a further optimisation of this strategy allowing a repositioning around the galaxies of interest. Starting from the galaxy targeting strategy described above, we proceed as follows to reposition a given tile:

\begin{enumerate}
    \item Initially the tile is centered on a galaxy of interest.
    \item Compute the list of all compatible galaxies that are within the tile or in its immediate vicinity (within $4$ times the FoV of MXT).
    \item Compute the barycenter of this list of galaxies, i.e. the barycenter of their positions weighter by their grade.
    \item If a tile centered on this barycenter (still having a size equal to the MXT FoV) contains all the galaxies in this list, keep this new tile and move to step (vi);
    \item else remove the galaxy in the list furthest from the barycenter and move again to step (iii).
    \item Check that the grade of the new tile is higher than the grade of the original one.
If not, keep the original tile of step (i). 
\end{enumerate}

\begin{figure*}
\begin{center}
\includegraphics[width=0.9\textwidth]{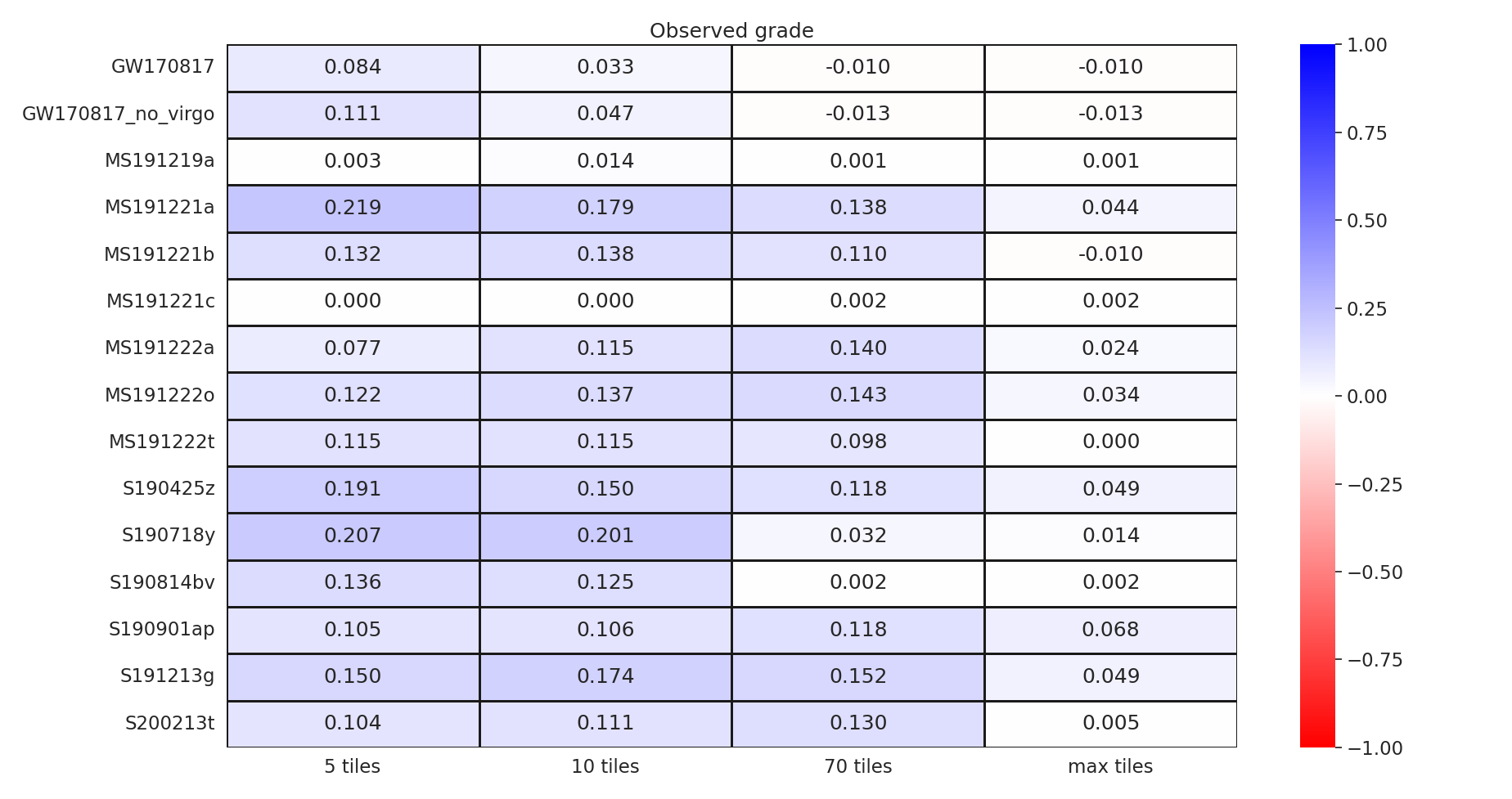}
\caption{\textbf{Comparison between the galaxy targeting strategy with and without including the repositioning procedure: observed grade.}
The difference 
between the observed grade $\mathcal{G}$
with the galaxy targeting strategy including the repositioning procedure defined in \S~\ref{subsection:reposition} and without including it
is given for each skymap listed in Table~\ref{tab:skymap}, assuming that  $N_{\mathrm{tiles}}=5$, $10$, $70$ and $N_{\mathrm{tiles,max}}$ tiles have already been observed.
This difference is normalized by the maximum observed grade $\mathcal{G}_{\mathrm{max}}$ (see Table~\ref{tab:naming} for the definition of these quantities).
The bluer the color, the better the galaxy targeting strategy implementing the repositioning procedure. Note that 70 tiles is the expected number of tiles allocated for a given ToO with SVOM.}
\label{fig:heatmap_grade_galaxy_wandering}
\end{center}
\end{figure*}

This repositioning procedure improves the flexibility of the galaxy targeting strategy and usually improves the grade of each tile, as illustrated in Figure~\ref{fig:three_skymaps_observed_grade} and \ref{fig:heatmap_grade_galaxy_wandering}.

\subsection{Optimisation for VT and MXT observations}
\label{subsection:optVT}

So far, we have only discussed MXT observations. However, the MXT shares its optical axis with the VT telescope, which has a smaller field of view. The implementation of the repositioning procedure described in the previous subsection raises the question of the optimization of the position of the compatible galaxies in the VT images. For this purpose, we implemented an additional step in the repositioning procedure:

\begin{enumerate}[resume]
    \item Check that at least one compatible galaxy is falling in the VT FoV. If not, keep the original tile of step (i).
\end{enumerate}
This ensures that every observation of a tile with the VT contributes to the search for a counterpart, as it contains at least a compatible galaxy.

\begin{figure*}
\begin{center}
\includegraphics[width=\textwidth]{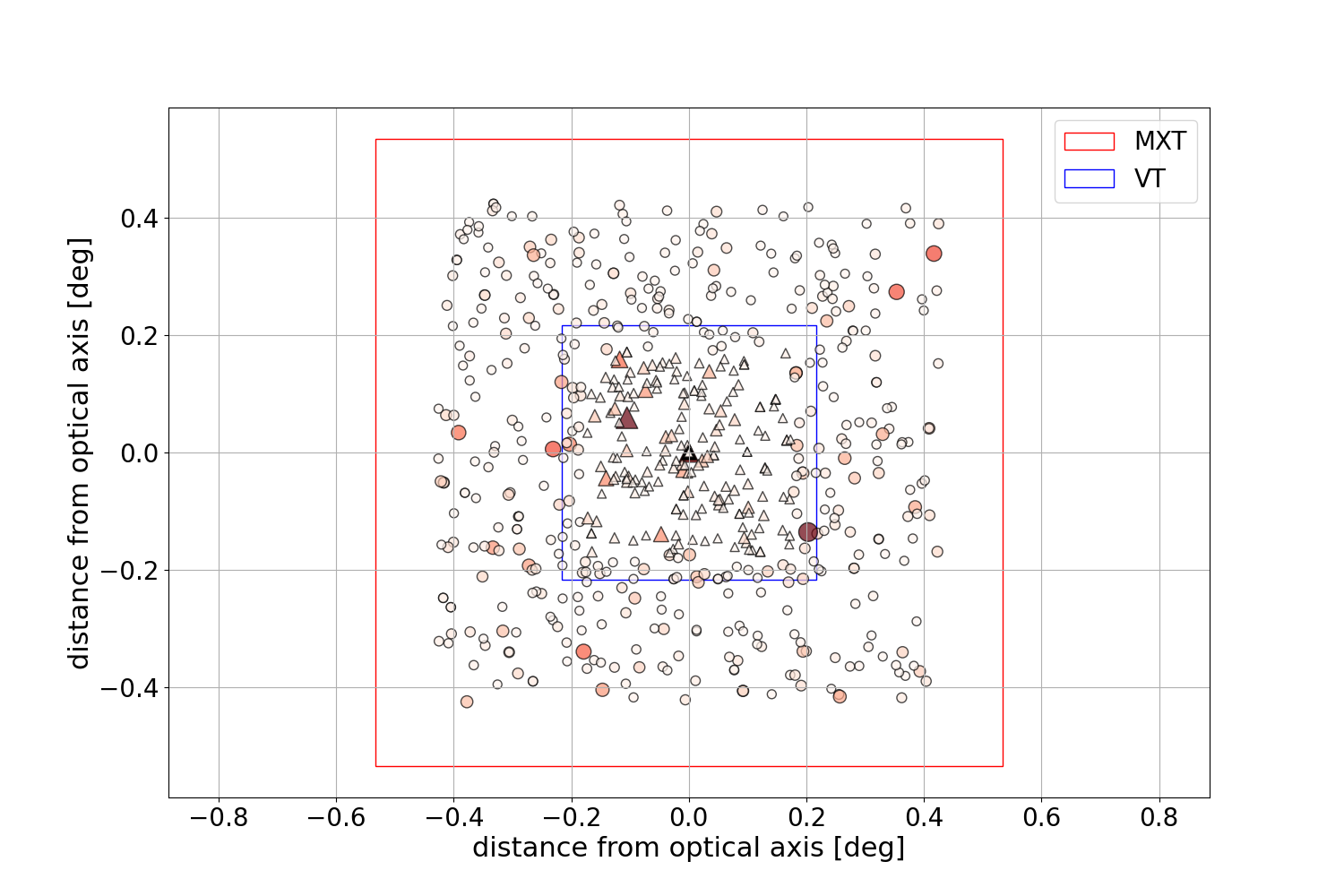}
\caption{
\textbf{Galaxy targeting strategy: edge bias.}
The relative position of all compatibles galaxies with respect to the optical axis is plotted in the MXT (red square) and VT (blue square) FoV for all accumalated tiles of the observation plan produced for the MS191222t skymap, including the repositioning procedure with the addition of the consideration of the bias effect, as described in Section~\ref{section:furtherdev}. Circles are galaxies considered as observed by MXT, and triangles are galaxies considered as observed by both  MXT and VT. The darker the colour, the higher the grade of the galaxy. The edge bias condition discussed in \S~\ref{subsection:optVT} is visible as an empty region at the edge of both MXT and VT FoV.}
\label{fig:FoV_MS191222t_galaxy_all_constraint}
\end{center}
\end{figure*}

We also considered the edge bias in MXT and VT images. While great efforts to ensure consistent localization performance in the entire focal plane, the position accuracy obtained by MXT starts to degrade in the 10\% of the image closest to the edge \citep{MXTloc}. Hence, we precautionary don't consider as observed a galaxy that falls in this edge region of MXT images, and we use the same criterion for the VT telescope. Figure \ref{fig:FoV_MS191222t_galaxy_all_constraint} shows an example of galaxy positions in the MXT and VT FoV for the observation plan generated for the MS191222t skymap using the galaxy targeting strategy with this edge bias condition implemented in the repositioning procedure. Note that, due to the conditions (vi) and (vii), there is a non-visible overlap of the galaxies close to the optical axis of MXT and VT.

\section{Discussion}
\label{section:TBD}

In this section, we discuss the compatibility of the proposed follow-up strategy with  the  expected properties of the electromagnetic counterparts to gravitational waves.

\subsection{Expected rate of GW events}
\label{subsection:Compatibility}
The galaxy targeting strategy developed in this work is limited to nearby LVK events where the galaxies catalogs can be used, i.e. distance where they have a reasonable completeness. As an illustration, the Mangrove catalog \citep{mangrove} is available up to 400 Mpc. \citet{O4expect} provide an expectation of about 20 BNS GW event per year below 400 Mpc, plus about 10 NSBH
 mergers \citep[see figure 2 in][]{O4expect}: note that only a sub-fraction of NSBH mergers are expected to be associated to electromagnetic counterparts, as it probably requires the tidal disruption of the neutron star before reaching the blak hole horizon. On the other hand, the expected rate of ToO for GW follow-up with SVOM is about one per month. Therefore, restricting the SVOM follow-up to these BNS and NSBH GW events below 400 Mpc during the runs O4 and O5 of LVK is compatible with the time allocated to ToO observations by the mission. As the expected rate of these GW events\footnote{which still suffers from great uncertainty \citep[see e.g.][]{O4expect}, which can only be reduced after the O4 run.}
is possibly slightly higher than the ToO rate defined in the current SVOM program, additional selection is kept possible to focus on the most promising events regarding the search for electromagnetic counterparts. 
Such selection could take into account the reliability of the GW source classification from LIGO-Virgo-KAGRA, the size of the skymap, the estimated distance of the source, etc.

\begin{figure}
\begin{center}
\includegraphics[width=1\columnwidth]{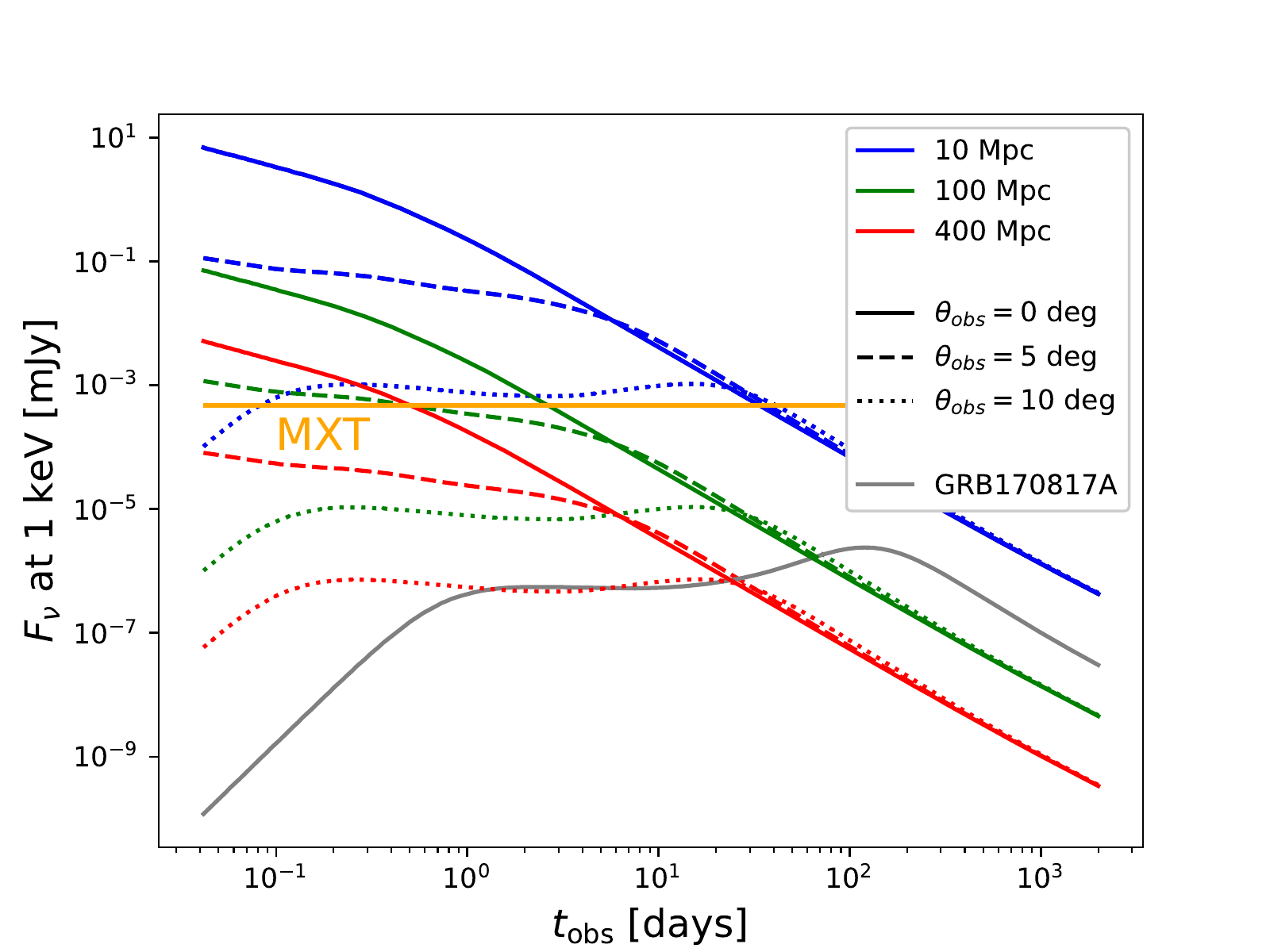}
\caption{Simulated afterglow lightcurves for GRB170817A like parameters with varying viewing angles $\theta_{\mathrm{obs}}$ and distances. The horizontal orange solid line shows
the expected limiting sensitivity for a 10 minutes long exposure with the MXT telescope. 
The grey solid line represent the lightcurve with best-fit parameters of the afterglow of GRB170817A ($\theta_{obs} \sim 21$ degree and distance $\sim 40$ Mpc), taken from \citet{pellouin}.}
\label{fig:170817_like}
\end{center}
\end{figure}

The case of a simultaneous detection of a GRB, which should remain rare during runs O4 and O5 \citep[see e.g.][]{2021A&A...651A..83M,O4expect,2022ApJ...928..186A}, will of course give the highest priority to a GW alert. Such a case may indicate an on-axis or slighltly off-axis observation, which is the most favorable case for the early detection of an X-ray counterpart, as discussed below (see also Figure~\ref{fig:170817_like}).
In the case of a GRB detection without precise localization, e.g. by Fermi/GBM or SVOM/GRM, the strategy described in this paper applies for the rapid follow-up by SVOM, in a favourable context thanks to a reduced error box. On the other hand, if the GRB is well localized, for example by Swift/BAT or SVOM/ECLAIRs, no tiling is required and the usual GRB follow-up strategy applies.

\subsection{Detectability of electromagnetic counterparts}

\subsubsection{MXT detectability}

The main electromagnetic counterpart expected to be detectable in X-rays by the MXT telescope in case of a BNS merger is the GRB afterglow. To illustrate the sensitivity of the MXT for such sources,
Figure~\ref{fig:170817_like} compares the predicted X-ray afterglow lightcurve of a GRB170817A-like afterglow at different distance and viewing angles with the typical MXT sensitivity for a 10 minutes exposure ($\sim 5 \times 10^{-4}$ mJy at 1 keV, \citealt{MXTperf}).
The afterglow lightcurves are simulated using the best-fit parameters of the GRB170817A afterglow with a detailed afterglow model of a relativistic structured jet. The model and  the best-fit parameters are taken from \citet{pellouin}, where a detailed description of the model is provided. Note the strong dependence on the viewing angle of the peak flux and the peak time, which is of course related to the ultra-relativistic nature of the jet.

Figure~\ref{fig:170817_like} shows that the MXT sensitivity is compatible with the expected flux of such a GRB afterglow at 400 Mpcif it is seen on-axis or only slightly off-axis.
For such viewing angles, the flux is the brightest at early times, hence the need for a rapid follow-up of GW alerts. This shows that any future improvement of the SVOM system to reduce the reaction delay between a GW alert and the first ToO observation will significantly increase the chance of the detection of an afterglow. It also highlights again the importance of optimizing the observations to explore first the regions where the probability of the GW source being present is highest.

Finally, Figure~\ref{fig:170817_like} also shows that more off-axis aferglows remain detectable by the MXT telescope at shorter distances. 
The peak of the lightcurve can be signficantly delayed in this case, which therefore requires a different strategy compared to the rapid follow-up discussed in this work. We leave the discussion of the best strategy  to implement for SVOM in such cases to a future paper.

\subsubsection{VT detectability}
\begin{figure}
\begin{center}
\includegraphics[width=1\columnwidth]{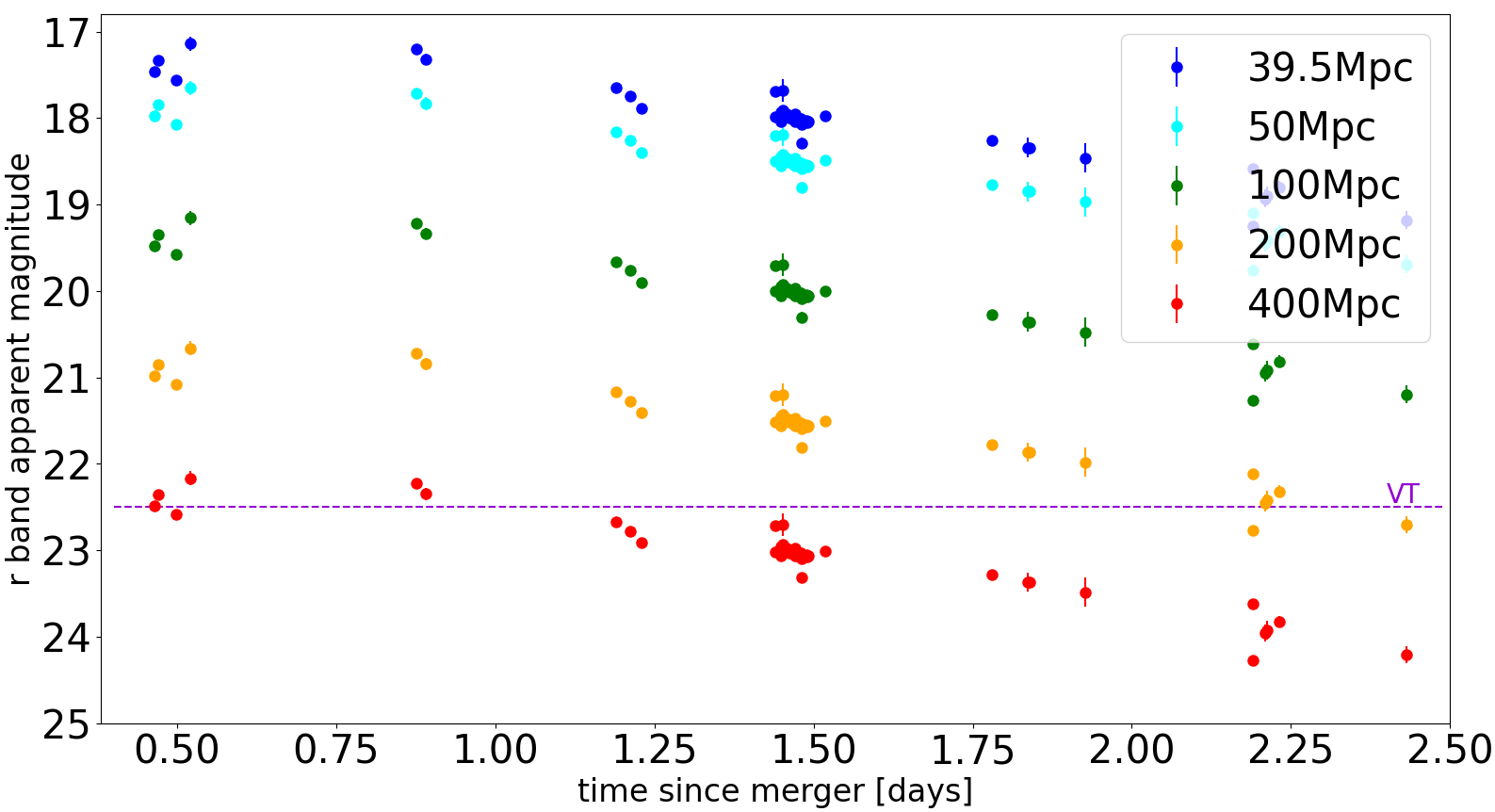}
\caption{Early light curve in r band of the observed kilonova associated to GW170817 (blue) and of a similar event at a larger distance of 50 (cyan), 100 (green), 200 (orange) and 400 Mpc (red). All data are taken from \citet{2017ApJ...851L..21V}. No additional K-correction is included. The horizontal purple dotted line shows the expected limiting magnitude for a 10 minutes exposure with the VT telescope.}
\label{fig:kilonova_VTlim}
\end{center}
\end{figure}

For the VT and other visible telescopes, the most promising electromagnetic counterpart is the expected kilonova emission. The absolute magnitude of a kilonova is relatively low: 
AT 2017gfo, the kilonova associated to GW 170817, peaked at an apparent magnitude of
$\sim$ 17 in r band (e.g. \citet{Arcavi2018}) despite a close distance of 40 Mpc. Figure~\ref{fig:kilonova_VTlim} shows the apparent magnitude in r band of a similar kilonova at distances ranging from 40 to 400 Mpc and expected limiting magnitude of 22.5 for a 10 minutes exposure time with the VT telescope. In agreement with \citet{2020Ap&SS.365..185A}, we find that the kilonova remains detectable by the VT up to 400 Mpc (the maximum distance imposed by the mangrove catalogue), even if the actual detection becomes challenging for the largest distances, the kilonova remaining above the VT sensitivity for only half a day at 400 Mpc, which stresses again the importance of a rapid follow-up.

If the relativistic jet is seen on-axis or slightly off-axis, the early afterglow is expected to be much brighter than the kilonova, as already observed in several cases such as in association with the short GRB 130603B \citep{2013Natur.500..547T}. As SVOM has been optimized for GRB observations, the VT sensitivity is well adapted to the detection of the visible afterglow in such a case \citep{SVOMwhitepaper}.

\section{Conclusions}
\label{section:conclusion}

The detection of an electromagnetic counterpart to a GW event is very challenging. The multi-wavelength capabilities of the SVOM mission are well adapted to this search. In this work we make use of recent developments in both catalogues of galaxies and galaxy targeting strategy to simulate and develop GW follow-up observation plan specific to the SVOM satellite. 
With the sensitivity of the second generation GW interefrometers  LVK, leading to an horizon of 400 Mpc for BNS mergers, we identify the galaxy targeting strategy as the most efficient strategy for the rapid follow-up in the X-ray and visible range with the MXT and VT telescopes onboard SVOM. We developed a realistic and optimised version of this galaxy targeting strategy, where several important constraints specific to the SVOM mission and the SVOM satellite platform are taken into account (exposure time, number of possible pointings during a ToO observation, slew limitation, edge bias in images, etc.) and where the tiling of the GW skymap and the ordered sequence of observations of the tiles are optimised to observe first the most promising host galaxy candidates in the regions where the probability of the GW source to be present is the highest. These observation plans are led by the MXT telescope but are also optimised for the best coverage of the compatible galaxies in the skymap by the VT telescope.

We also checked that the detection rate of BNS or NSBH merger expected for LVK in the coming observing runs and within a distance of 400 Mpc imposed by the use of galaxy catalogs is compatible with the time allocated by the SVOM mission for the rapid follow-up of such events. We also showed that the sensitivity of the MXT and VT telescopes allows the detection of some expected electromagnetic counterparts within the same distance: kilonova, and visible and X-ray afterglows when the relativistic jet is seen on-axis or slightly offaxis. Some offaxis afterglows may also be detectable by these two instruments at lower distance, but will require a specific follow-up strategy as their flux can reach its peak with a long delay from the GW signal. This case will be discussed in a future paper.  
 
Developments presented in this work are essential for the production of realistically optimised observation plans that SVOM will use in the near future for the rapid follow-up of GW events.

\section*{Acknowledgements}

The authors acknowledge the Centre National d’Études Spatiales (CNES) for financial support in this research project. This project was supported by a research grant from the Ile-de-France Region within the framework of the Domaine d’Intérêt Majeur-Astrophysique et Conditions d’Apparition de la Vie (DIM-ACAV). This work has made use of the Infinity Cluster hosted by Institut d'Astrophysique de Paris.

\section*{Data availability}

No new data were generated or analysed in support of this research.



\bibliographystyle{mnras}
\bibliography{refs} 

\begin{thebibliography}{}
\makeatletter
\relax
\def\mn@urlcharsother{\let\do\@makeother \do\$\do\&\do\#\do\^\do\_\do\%\do\~}
\def\mn@doi{\begingroup\mn@urlcharsother \@ifnextchar [ {\mn@doi@}
  {\mn@doi@[]}}
\def\mn@doi@[#1]#2{\def\@tempa{#1}\ifx\@tempa\@empty \href
  {http://dx.doi.org/#2} {doi:#2}\else \href {http://dx.doi.org/#2} {#1}\fi
  \endgroup}
\def\mn@eprint#1#2{\mn@eprint@#1:#2::\@nil}
\def\mn@eprint@arXiv#1{\href {http://arxiv.org/abs/#1} {{\tt arXiv:#1}}}
\def\mn@eprint@dblp#1{\href {http://dblp.uni-trier.de/rec/bibtex/#1.xml}
  {dblp:#1}}
\def\mn@eprint@#1:#2:#3:#4\@nil{\def\@tempa {#1}\def\@tempb {#2}\def\@tempc
  {#3}\ifx \@tempc \@empty \let \@tempc \@tempb \let \@tempb \@tempa \fi \ifx
  \@tempb \@empty \def\@tempb {arXiv}\fi \@ifundefined
  {mn@eprint@\@tempb}{\@tempb:\@tempc}{\expandafter \expandafter \csname
  mn@eprint@\@tempb\endcsname \expandafter{\@tempc}}}

\bibitem[\protect\citeauthoryear{Abbott et~al.}{Abbott
  et~al.}{2017a}]{LSC_BNS_2017PhRvL}
Abbott B.~P.,  et~al., 2017a, \mn@doi [Phys. Rev. Lett.]
  {10.1103/PhysRevLett.119.161101}, 119, 161101

\bibitem[\protect\citeauthoryear{{Abbott} et~al.,}{{Abbott}
  et~al.}{2017b}]{gwtohubble1}
{Abbott} B.~P.,  et~al., 2017b, \mn@doi [\nat] {10.1038/nature24471}, \href
  {https://ui.adsabs.harvard.edu/abs/2017Natur.551...85A} {551, 85}

\bibitem[\protect\citeauthoryear{{Abbott} et~al.,}{{Abbott}
  et~al.}{2017c}]{2017ApJ...848L..13A}
{Abbott} B.~P.,  et~al., 2017c, \mn@doi [\apjl] {10.3847/2041-8213/aa920c},
  \href {https://ui.adsabs.harvard.edu/abs/2017ApJ...848L..13A} {848, L13}

\bibitem[\protect\citeauthoryear{{Abbott} et~al.,}{{Abbott}
  et~al.}{2018}]{2018PhRvL.121p1101A}
{Abbott} B.~P.,  et~al., 2018, \mn@doi [\prl] {10.1103/PhysRevLett.121.161101},
  \href {https://ui.adsabs.harvard.edu/abs/2018PhRvL.121p1101A} {121, 161101}

\bibitem[\protect\citeauthoryear{{Abbott} et~al.,}{{Abbott}
  et~al.}{2022}]{2022ApJ...928..186A}
{Abbott} R.,  et~al., 2022, \mn@doi [\apj] {10.3847/1538-4357/ac532b}, \href
  {https://ui.adsabs.harvard.edu/abs/2022ApJ...928..186A} {928, 186}

\bibitem[\protect\citeauthoryear{{Antolini}, {Caiazzo}, {Dav{\'e}}  \&
  {Heyl}}{{Antolini} et~al.}{2017}]{Antolini2017}
{Antolini} E.,  {Caiazzo} I.,  {Dav{\'e}} R.,   {Heyl} J.~S.,  2017, \mn@doi
  [\mnras] {10.1093/mnras/stw3292}, \href
  {https://ui.adsabs.harvard.edu/abs/2017MNRAS.466.2212A} {466, 2212}

\bibitem[\protect\citeauthoryear{{Arcavi}}{{Arcavi}}{2018}]{Arcavi2018}
{Arcavi} I.,  2018, \mn@doi [\apjl] {10.3847/2041-8213/aab267}, \href
  {https://ui.adsabs.harvard.edu/abs/2018ApJ...855L..23A} {855, L23}

\bibitem[\protect\citeauthoryear{{Arcavi} et~al.,}{{Arcavi}
  et~al.}{2017}]{LosC}
{Arcavi} I.,  et~al., 2017, \mn@doi [\apjl] {10.3847/2041-8213/aa910f}, \href
  {https://ui.adsabs.harvard.edu/abs/2017ApJ...848L..33A} {848, L33}

\bibitem[\protect\citeauthoryear{{Arcier} et~al.,}{{Arcier}
  et~al.}{2020}]{2020Ap&SS.365..185A}
{Arcier} B.,  et~al., 2020, \mn@doi [\apss] {10.1007/s10509-020-03898-z}, \href
  {https://ui.adsabs.harvard.edu/abs/2020Ap&SS.365..185A} {365, 185}

\bibitem[\protect\citeauthoryear{{Artale}, {Mapelli}, {Giacobbo}, {Sabha},
  {Spera}, {Santoliquido}  \& {Bressan}}{{Artale}
  et~al.}{2019}]{2019MNRAS.487.1675A}
{Artale} M.~C.,  {Mapelli} M.,  {Giacobbo} N.,  {Sabha} N.~B.,  {Spera} M.,
  {Santoliquido} F.,   {Bressan} A.,  2019, \mn@doi [\mnras]
  {10.1093/mnras/stz1382}, \href
  {https://ui.adsabs.harvard.edu/abs/2019MNRAS.487.1675A} {487, 1675}

\bibitem[\protect\citeauthoryear{{Berger}}{{Berger}}{2014}]{Berger2014}
{Berger} E.,  2014, \mn@doi [\araa] {10.1146/annurev-astro-081913-035926},
  \href {https://ui.adsabs.harvard.edu/abs/2014ARA&A..52...43B} {52, 43}

\bibitem[\protect\citeauthoryear{{Cantiello} et~al.,}{{Cantiello}
  et~al.}{2018}]{Cantiello2018}
{Cantiello} M.,  et~al., 2018, \mn@doi [\apjl] {10.3847/2041-8213/aaad64},
  \href {https://ui.adsabs.harvard.edu/abs/2018ApJ...854L..31C} {854, L31}

\bibitem[\protect\citeauthoryear{{Cook} et~al.,}{{Cook}
  et~al.}{2019}]{2019ApJ...880....7C}
{Cook} D.~O.,  et~al., 2019, \mn@doi [\apj] {10.3847/1538-4357/ab2131}, \href
  {https://ui.adsabs.harvard.edu/abs/2019ApJ...880....7C} {880, 7}

\bibitem[\protect\citeauthoryear{{Coughlin} et~al.,}{{Coughlin}
  et~al.}{2018}]{gwemopt}
{Coughlin} M.~W.,  et~al., 2018, \mn@doi [\mnras] {10.1093/mnras/sty1066},
  \href {https://ui.adsabs.harvard.edu/abs/2018MNRAS.478..692C} {478, 692}

\bibitem[\protect\citeauthoryear{{Coughlin}, {Dietrich}, {Heinzel}, {Khetan},
  {Antier}, {Christensen}, {Coulter}  \& {Foley}}{{Coughlin}
  et~al.}{2019a}]{gwtohubble2}
{Coughlin} M.~W.,  {Dietrich} T.,  {Heinzel} J.,  {Khetan} N.,  {Antier} S.,
  {Christensen} N.,  {Coulter} D.~A.,   {Foley} R.~J.,  2019a, arXiv e-prints,
  \href {https://ui.adsabs.harvard.edu/abs/2019arXiv190800889C} {p.
  arXiv:1908.00889}

\bibitem[\protect\citeauthoryear{{Coughlin} et~al.,}{{Coughlin}
  et~al.}{2019b}]{Coughlin19_opt}
{Coughlin} M.~W.,  et~al., 2019b, arXiv e-prints, \href
  {https://ui.adsabs.harvard.edu/abs/2019arXiv190901244C} {p. arXiv:1909.01244}

\bibitem[\protect\citeauthoryear{{D'Avanzo} et~al.,}{{D'Avanzo}
  et~al.}{2018}]{DAvanzo2018}
{D'Avanzo} P.,  et~al., 2018, \mn@doi [\aap] {10.1051/0004-6361/201832664},
  \href {https://ui.adsabs.harvard.edu/abs/2018A&A...613L...1D} {613, L1}

\bibitem[\protect\citeauthoryear{{D{\'a}lya} et~al.,}{{D{\'a}lya}
  et~al.}{2018}]{2018MNRAS.479.2374D}
{D{\'a}lya} G.,  et~al., 2018, \mn@doi [\mnras] {10.1093/mnras/sty1703}, \href
  {https://ui.adsabs.harvard.edu/abs/2018MNRAS.479.2374D} {479, 2374}

\bibitem[\protect\citeauthoryear{{D{\'a}lya} et~al.,}{{D{\'a}lya}
  et~al.}{2022}]{glade+}
{D{\'a}lya} G.,  et~al., 2022, \mn@doi [\mnras] {10.1093/mnras/stac1443}, \href
  {https://ui.adsabs.harvard.edu/abs/2022MNRAS.514.1403D} {514, 1403}

\bibitem[\protect\citeauthoryear{{Dong}, {Wu}, {Li}, {Zhang}  \&
  {Zhang}}{{Dong} et~al.}{2010}]{GRM}
{Dong} Y.,  {Wu} B.,  {Li} Y.,  {Zhang} Y.,   {Zhang} S.,  2010, \mn@doi
  [Science China Physics, Mechanics, and Astronomy]
  {10.1007/s11433-010-0011-7}, \href
  {https://ui.adsabs.harvard.edu/abs/2010SCPMA..53S..40D} {53, 40}

\bibitem[\protect\citeauthoryear{{Ducoin}, {Corre}, {Leroy}  \& {Le
  Floch}}{{Ducoin} et~al.}{2020}]{mangrove}
{Ducoin} J.~G.,  {Corre} D.,  {Leroy} N.,   {Le Floch} E.,  2020, \mn@doi
  [\mnras] {10.1093/mnras/staa114}, \href
  {https://ui.adsabs.harvard.edu/abs/2020MNRAS.492.4768D} {492, 4768}

\bibitem[\protect\citeauthoryear{{Ebrov{\'a}}, {B{\'\i}lek}, {Y{\i}ld{\i}z}  \&
  {Eli{\'a}{\v{s}}ek}}{{Ebrov{\'a}} et~al.}{2020}]{Ebrova2020}
{Ebrov{\'a}} I.,  {B{\'\i}lek} M.,  {Y{\i}ld{\i}z} M.~K.,   {Eli{\'a}{\v{s}}ek}
  J.,  2020, \mn@doi [\aap] {10.1051/0004-6361/201935219}, \href
  {https://ui.adsabs.harvard.edu/abs/2020A&A...634A..73E} {634, A73}

\bibitem[\protect\citeauthoryear{{Fan} et~al.,}{{Fan} et~al.}{2020}]{VT}
{Fan} X.,  et~al., 2020, in Society of Photo-Optical Instrumentation Engineers
  (SPIE) Conference Series. p. 114430Q, \mn@doi{10.1117/12.2561854}

\bibitem[\protect\citeauthoryear{{Fong} et~al.,}{{Fong}
  et~al.}{2013}]{Fong2013}
{Fong} W.,  et~al., 2013, \mn@doi [\apj] {10.1088/0004-637X/769/1/56}, \href
  {https://ui.adsabs.harvard.edu/abs/2013ApJ...769...56F} {769, 56}

\bibitem[\protect\citeauthoryear{{Fong} et~al.,}{{Fong}
  et~al.}{2022}]{2022arXiv220601763F}
{Fong} W.-f.,  et~al., 2022, arXiv e-prints, \href
  {https://ui.adsabs.harvard.edu/abs/2022arXiv220601763F} {p. arXiv:2206.01763}

\bibitem[\protect\citeauthoryear{{Gehrels}, {Cannizzo}, {Kanner}, {Kasliwal},
  {Nissanke}  \& {Singer}}{{Gehrels} et~al.}{2016}]{2016ApJ...820..136G}
{Gehrels} N.,  {Cannizzo} J.~K.,  {Kanner} J.,  {Kasliwal} M.~M.,  {Nissanke}
  S.,   {Singer} L.~P.,  2016, \mn@doi [\apj] {10.3847/0004-637X/820/2/136},
  \href {https://ui.adsabs.harvard.edu/abs/2016ApJ...820..136G} {820, 136}

\bibitem[\protect\citeauthoryear{{Ghosh}, {Bloemen}, {Nelemans}, {Groot}  \&
  {Price}}{{Ghosh} et~al.}{2016}]{2016A&A...592A..82G}
{Ghosh} S.,  {Bloemen} S.,  {Nelemans} G.,  {Groot} P.~J.,   {Price} L.~R.,
  2016, \mn@doi [\aap] {10.1051/0004-6361/201527712}, \href
  {https://ui.adsabs.harvard.edu/abs/2016A&A...592A..82G} {592, A82}

\bibitem[\protect\citeauthoryear{{Gill}, {Nathanail}  \& {Rezzolla}}{{Gill}
  et~al.}{2019}]{2019ApJ...876..139G}
{Gill} R.,  {Nathanail} A.,   {Rezzolla} L.,  2019, \mn@doi [\apj]
  {10.3847/1538-4357/ab16da}, \href
  {https://ui.adsabs.harvard.edu/abs/2019ApJ...876..139G} {876, 139}

\bibitem[\protect\citeauthoryear{{G{\'o}rski}, {Hivon}, {Banday}, {Wandelt},
  {Hansen}, {Reinecke}  \& {Bartelmann}}{{G{\'o}rski} et~al.}{2005}]{HEALPix}
{G{\'o}rski} K.~M.,  {Hivon} E.,  {Banday} A.~J.,  {Wandelt} B.~D.,  {Hansen}
  F.~K.,  {Reinecke} M.,   {Bartelmann} M.,  2005, \mn@doi [\apj]
  {10.1086/427976}, \href
  {https://ui.adsabs.harvard.edu/abs/2005ApJ...622..759G} {622, 759}

\bibitem[\protect\citeauthoryear{{Gotz} et~al.,}{{Gotz} et~al.}{2015}]{MXT}
{Gotz} D.,  et~al., 2015, arXiv e-prints, \href
  {https://ui.adsabs.harvard.edu/abs/2015arXiv150700204G} {p. arXiv:1507.00204}

\bibitem[\protect\citeauthoryear{{Gotz} et~al.,}{{Gotz} et~al.}{2022}]{MXTperf}
{Gotz} D.,  et~al., 2022, arXiv e-prints, \href
  {https://ui.adsabs.harvard.edu/abs/2022arXiv221113489G} {p. arXiv:2211.13489}

\bibitem[\protect\citeauthoryear{{Hajela} et~al.,}{{Hajela}
  et~al.}{2019a}]{2019arXiv190906393H}
{Hajela} A.,  et~al., 2019a, arXiv e-prints, \href
  {https://ui.adsabs.harvard.edu/abs/2019arXiv190906393H} {p. arXiv:1909.06393}

\bibitem[\protect\citeauthoryear{{Hajela} et~al.,}{{Hajela}
  et~al.}{2019b}]{Hajela2019}
{Hajela} A.,  et~al., 2019b, \mn@doi [\apjl] {10.3847/2041-8213/ab5226}, \href
  {https://ui.adsabs.harvard.edu/abs/2019ApJ...886L..17H} {886, L17}

\bibitem[\protect\citeauthoryear{{Hotokezaka}, {Nakar}, {Gottlieb}, {Nissanke},
  {Masuda}, {Hallinan}, {Mooley}  \& {Deller}}{{Hotokezaka}
  et~al.}{2019}]{gwtohubble3}
{Hotokezaka} K.,  {Nakar} E.,  {Gottlieb} O.,  {Nissanke} S.,  {Masuda} K.,
  {Hallinan} G.,  {Mooley} K.~P.,   {Deller} A.~T.,  2019, \mn@doi [Nature
  Astronomy] {10.1038/s41550-019-0820-1}, \href
  {https://ui.adsabs.harvard.edu/abs/2019NatAs...3..940H} {3, 940}

\bibitem[\protect\citeauthoryear{{Hussein}, {Robinet}, {Boutelier}, {G{\"o}tz},
  {Gros}  \& {Schneider}}{{Hussein} et~al.}{2022}]{MXTloc}
{Hussein} S.,  {Robinet} F.,  {Boutelier} M.,  {G{\"o}tz} D.,  {Gros} A.,
  {Schneider} B.,  2022, arXiv e-prints, \href
  {https://ui.adsabs.harvard.edu/abs/2022arXiv220913330H} {p. arXiv:2209.13330}

\bibitem[\protect\citeauthoryear{{Leibler} \& {Berger}}{{Leibler} \&
  {Berger}}{2010}]{Leibler2010}
{Leibler} C.~N.,  {Berger} E.,  2010, \mn@doi [\apj]
  {10.1088/0004-637X/725/1/1202}, \href
  {https://ui.adsabs.harvard.edu/abs/2010ApJ...725.1202L} {725, 1202}

\bibitem[\protect\citeauthoryear{Li, Mao, Qin, Zheng, Liu, Zhao  \& Zhao}{Li
  et~al.}{2023}]{Yubin2023}
Li Y.,  Mao J.,  Qin J.,  Zheng X.,  Liu F.,  Zhao Y.,   Zhao X.-H.,  2023,
  Research in Astronomy and Astrophysics

\bibitem[\protect\citeauthoryear{{Mapelli}, {Giacobbo}, {Toffano}, {Ripamonti},
  {Bressan}, {Spera}  \& {Branchesi}}{{Mapelli}
  et~al.}{2018}]{2018MNRAS.481.5324M}
{Mapelli} M.,  {Giacobbo} N.,  {Toffano} M.,  {Ripamonti} E.,  {Bressan} A.,
  {Spera} M.,   {Branchesi} M.,  2018, \mn@doi [\mnras]
  {10.1093/mnras/sty2663}, \href
  {https://ui.adsabs.harvard.edu/abs/2018MNRAS.481.5324M} {481, 5324}

\bibitem[\protect\citeauthoryear{{Metzger}}{{Metzger}}{2019}]{Metzgerkilo}
{Metzger} B.~D.,  2019, arXiv e-prints, \href
  {https://ui.adsabs.harvard.edu/abs/2019arXiv191001617M} {p. arXiv:1910.01617}

\bibitem[\protect\citeauthoryear{{Mochkovitch}, {Daigne}, {Duque}  \&
  {Zitouni}}{{Mochkovitch} et~al.}{2021}]{2021A&A...651A..83M}
{Mochkovitch} R.,  {Daigne} F.,  {Duque} R.,   {Zitouni} H.,  2021, \mn@doi
  [\aap] {10.1051/0004-6361/202140689}, \href
  {https://ui.adsabs.harvard.edu/abs/2021A&A...651A..83M} {651, A83}

\bibitem[\protect\citeauthoryear{{Nugent} et~al.,}{{Nugent}
  et~al.}{2022}]{2022arXiv220601764N}
{Nugent} A.~E.,  et~al., 2022, arXiv e-prints, \href
  {https://ui.adsabs.harvard.edu/abs/2022arXiv220601764N} {p. arXiv:2206.01764}

\bibitem[\protect\citeauthoryear{{Pellouin} \& {Daigne}}{{Pellouin} \&
  {Daigne}}{2023}]{pellouin}
{Pellouin} C.,  {Daigne} F.,  2023, in preparation

\bibitem[\protect\citeauthoryear{{Petrov} et~al.,}{{Petrov}
  et~al.}{2022}]{O4expect}
{Petrov} P.,  et~al., 2022, \mn@doi [\apj] {10.3847/1538-4357/ac366d}, \href
  {https://ui.adsabs.harvard.edu/abs/2022ApJ...924...54P} {924, 54}

\bibitem[\protect\citeauthoryear{{Planck Collaboration}}{{Planck
  Collaboration}}{2016}]{planck15}
{Planck Collaboration} 2016, \mn@doi [\aap] {10.1051/0004-6361/201525830},
  \href {https://ui.adsabs.harvard.edu/abs/2016A&A...594A..13P} {594, A13}

\bibitem[\protect\citeauthoryear{{Rana} \& {Mooley}}{{Rana} \&
  {Mooley}}{2019}]{Rana2019}
{Rana} J.,  {Mooley} K.~P.,  2019, arXiv e-prints, \href
  {https://ui.adsabs.harvard.edu/abs/2019arXiv190407335R} {p. arXiv:1904.07335}

\bibitem[\protect\citeauthoryear{{Santoliquido}, {Mapelli}, {Artale}  \&
  {Boco}}{{Santoliquido} et~al.}{2022}]{2022arXiv220505099S}
{Santoliquido} F.,  {Mapelli} M.,  {Artale} M.~C.,   {Boco} L.,  2022, arXiv
  e-prints, \href {https://ui.adsabs.harvard.edu/abs/2022arXiv220505099S} {p.
  arXiv:2205.05099}

\bibitem[\protect\citeauthoryear{{Schanne}, {Cordier}, {Atteia}, {Godet},
  {Lachaud}  \& {Mercier}}{{Schanne} et~al.}{2015}]{ECLAIRs}
{Schanne} S.,  {Cordier} B.,  {Atteia} J.-L.,  {Godet} O.,  {Lachaud} C.,
  {Mercier} K.,  2015, arXiv e-prints, \href
  {https://ui.adsabs.harvard.edu/abs/2015arXiv150805851S} {p. arXiv:1508.05851}

\bibitem[\protect\citeauthoryear{{Singer} \& {Price}}{{Singer} \&
  {Price}}{2016}]{bayestar}
{Singer} L.~P.,  {Price} L.~R.,  2016, \mn@doi [\prd]
  {10.1103/PhysRevD.93.024013}, \href
  {https://ui.adsabs.harvard.edu/abs/2016PhRvD..93b4013S} {93, 024013}

\bibitem[\protect\citeauthoryear{{Tanvir}, {Levan}, {Fruchter}, {Hjorth},
  {Hounsell}, {Wiersema}  \& {Tunnicliffe}}{{Tanvir}
  et~al.}{2013}]{2013Natur.500..547T}
{Tanvir} N.~R.,  {Levan} A.~J.,  {Fruchter} A.~S.,  {Hjorth} J.,  {Hounsell}
  R.~A.,  {Wiersema} K.,   {Tunnicliffe} R.~L.,  2013, \mn@doi [\nat]
  {10.1038/nature12505}, \href
  {https://ui.adsabs.harvard.edu/abs/2013Natur.500..547T} {500, 547}

\bibitem[\protect\citeauthoryear{{The LIGO Scientific Collaboration}
  et~al.,}{{The LIGO Scientific Collaboration} et~al.}{2021a}]{gwtc2.1}
{The LIGO Scientific Collaboration} et~al., 2021a, arXiv e-prints, \href
  {https://ui.adsabs.harvard.edu/abs/2021arXiv210801045T} {p. arXiv:2108.01045}

\bibitem[\protect\citeauthoryear{{The LIGO Scientific Collaboration}
  et~al.,}{{The LIGO Scientific Collaboration} et~al.}{2021b}]{gwtc3}
{The LIGO Scientific Collaboration} et~al., 2021b, arXiv e-prints, \href
  {https://ui.adsabs.harvard.edu/abs/2021arXiv211103606T} {p. arXiv:2111.03606}

\bibitem[\protect\citeauthoryear{{Toffano}, {Mapelli}, {Giacobbo}, {Artale}  \&
  {Ghirlanda}}{{Toffano} et~al.}{2019}]{2019MNRAS.tmp.2085T}
{Toffano} M.,  {Mapelli} M.,  {Giacobbo} N.,  {Artale} M.~C.,   {Ghirlanda} G.,
   2019, \mn@doi [\mnras] {10.1093/mnras/stz2415}, \href
  {https://ui.adsabs.harvard.edu/abs/2019MNRAS.tmp.2085T} {p.~2085}

\bibitem[\protect\citeauthoryear{{Villar} et~al.,}{{Villar}
  et~al.}{2017a}]{Villar2017}
{Villar} V.~A.,  et~al., 2017a, \mn@doi [\apjl] {10.3847/2041-8213/aa9c84},
  \href {https://ui.adsabs.harvard.edu/abs/2017ApJ...851L..21V} {851, L21}

\bibitem[\protect\citeauthoryear{{Villar et al.}}{{Villar et
  al.}}{2017b}]{2017ApJ...851L..21V}
{Villar et al.} V.~A.,  2017b, \mn@doi [Astrophysical Journall]
  {10.3847/2041-8213/aa9c84}, \href
  {https://ui.adsabs.harvard.edu/abs/2017ApJ...851L..21V} {851, L21}

\bibitem[\protect\citeauthoryear{{Wei et al.}}{{Wei et al.}}{2016a}]{SVOM2016}
{Wei et al.} J.,  2016a, arXiv e-prints, \href
  {https://ui.adsabs.harvard.edu/abs/2016arXiv161006892W} {p. arXiv:1610.06892}

\bibitem[\protect\citeauthoryear{{Wei} et~al.,}{{Wei}
  et~al.}{2016b}]{SVOMwhitepaper}
{Wei} J.,  et~al., 2016b, arXiv e-prints, \href
  {https://ui.adsabs.harvard.edu/abs/2016arXiv161006892W} {p. arXiv:1610.06892}

\makeatother
\end{thebibliography}







\bsp	
\label{lastpage}
\end{document}